\documentclass[doc,12pt]{apa}

\usepackage{apacite}
\usepackage[USenglish]{babel}
\usepackage[utf8x]{inputenc}
\usepackage{amsmath}
\usepackage{graphicx}
\usepackage[colorinlistoftodos]{todonotes}
\usepackage[pagewise]{lineno}
\usepackage{setspace}

\title{\LARGE\bf How spatial frequencies and color drive object search in real-world scenes:\\ [1.5ex]A new eye-movement corpus} 
\author{Anke Cajar, Ralf Engbert, \& Jochen Laubrock}
\vspace{1cm}
\affiliation{\large{University of Potsdam, Germany}}
\rightheader{Scene search with central and peripheral filters}

\note{\begin{flushleft} 
\vspace{8cm}
\normalsize{Address correspondence to:\\
Anke Cajar\\
Department Psychologie\\
Universit\"at Potsdam\\
Karl-Liebknecht-Str. 24-25\\
14476 Potsdam, Germany\\
e-mail: cajar@uni-potsdam.de\\
phone: +49-331-9772872\\
fax: +49-331-9772793}\end{flushleft}}
\newpage

\abstract{\normalsize{When studying how people search for objects in scenes, the inhomogeneity of the visual field is often ignored. Due to physiological limitations peripheral vision is blurred and mainly uses coarse-grained information (i.e., low spatial frequencies) for selecting saccade targets, whereas high-acuity central vision uses fine-grained information (i.e., high spatial frequencies) for analysis of details. Here we investigated how spatial frequencies and color affect object search in real-world scenes. Using gaze-contingent filters we attenuated high or low frequencies in central or peripheral vision while viewers searched color or grayscale scenes. Results showed that peripheral filters and central high-pass filters hardly affected search accuracy, whereas accuracy dropped drastically with central low-pass filters. Peripheral filtering increased the time to localize the target by decreasing saccade amplitudes and increasing number and duration of fixations. The use of coarse-grained information in the periphery was limited to color scenes. Central filtering increased the time to verify target identity instead, especially with low-pass filters. We conclude that peripheral vision is critical for object localization and central vision is critical for object identification. Visual guidance during peripheral object localization is dominated by low-frequency color information, whereas high-frequency information, relatively independent of color, is most important for object identification in central vision.}} 

\begin{document}

\thispagestyle{empty}
\maketitle

\newpage

\section{Introduction}

Searching for an object in a visual scene is a vital task we perform countless times in our daily lives. During search, we make about three eye movements per second called saccades, which rapidly shift our gaze to new points of interest in the scene. This is necessary because high acuity vision is only achieved in the fovea, the central 2$^\circ$ around fixation; toward the visual periphery, resolution falls off rapidly~\cite{Jones.JOptSocAm.1947,Wertheim.ZPsycholPhysiolSinnesorgane.1894}. Thus, fine-grained scene information, which is carried by high spatial frequencies, is processed best in central vision (from the point of fixation up to about 4-5$^\circ$ eccentricity, cf.~\citeNP{Larson.JVis.2009}). Central vision is necessary for analyzing details in a scene, identifying objects \cite{Henderson.PsycholSci.1999,Henderson.PerceptPsychophys.2003}, and establishing object memory \cite{Geringswald.JVis.2016}. Low-resolution peripheral vision, on the other hand, is best suited for processing coarse-grained information, which is carried by low spatial frequencies. Peripheral vision is mainly used for rapid reorienting and selecting new regions of interest as saccade targets. Central and peripheral vision can therefore be considered to serve different tasks \cite{Gilchrist.INBOOK.2011}. However, there are not many studies directly testing the consequences of the inhomogeneity of the visual field for scene perception, and not much is known about the different contributions of central and peripheral vision to object search in real-world scenes \cite<but see, e.g.,>{Nuthmann.VisCogn.2013,Nuthmann.JExpPsycholHuman.2014,Nuthmann.JVis.2016}.

The present study investigates central and peripheral vision during object-in-scene search. We were interested in whether the two parts of the visual field differ in the use of fine- and coarse-grained information, and whether color or brightness contrasts modulate its use. To this end, we attenuated high or low spatial frequencies in central or peripheral vision during search using gaze-contingent low-pass or high-pass filters respectively. Given the different sensitivities of the central and peripheral visual field to certain spatial-frequency bands~\cite<e.g.,>{Hilz.VisionRes.1974}, differences in search performance and eye-movement behavior can be expected depending on which frequencies are filtered in central and peripheral vision.

Previous studies show that saccade target selection during scene viewing is modulated by the available spatial-frequency information across the visual field: Saccades preferentially target unfiltered scene regions, as peripheral filtering decreases and central filtering increases saccade amplitudes~ \cite{Cajar.VisionRes.2016,Cajar.JVis.2016,Foulsham.AttenPerceptPsychophys.2011,Laubrock.JVis.2013,Loschky.JExpPsycholAppl.2002,Loschky.VisCogn.2005,Nuthmann.VisCogn.2013,Nuthmann.JExpPsycholHuman.2014}. These saccadic changes have been shown to reflect attentional modulations: Peripheral filtering induces tunnel vision with a narrowing of the attentional focus, and central filtering induces an attentional bias toward the periphery~\cite{Cajar.JVis.2016}. In scene memorization tasks, these saccadic modulations were larger when spatial frequencies that the respective part of the visual field is most sensitive to were attenuated, that is, with central low-pass and peripheral high-pass filtering~\cite{Cajar.VisionRes.2016,Cajar.JVis.2016,Laubrock.JVis.2013}. Fixation durations, on the other hand, did not increase as much or even decreased compared with unfiltered scene viewing in these conditions. Durations increased more with central high-pass and peripheral low-pass filtering, which preserve critical spatial frequencies, suggesting that fixation durations only prolong when the available information can be processed in a reasonable amount of time~\cite{Cajar.VisionRes.2016,Cajar.JVis.2016,Laubrock.JVis.2013}. This behavior can be explained by a computational model for fixation durations during scene viewing that assumes parallel processing of central and peripheral information during fixation with dynamical interactions between central and peripheral processing~\cite{Laubrock.JVis.2013}.

These previous findings imply greater scene processing difficulty when high spatial frequencies are attenuated in central vision and low spatial frequencies are attenuated in peripheral vision. The results derive from scene memorization tasks, however, with hardly any effects of filter type on task performance~\cite{Cajar.VisionRes.2016,Laubrock.JVis.2013}. Although we assume that viewers adjust their fixation duration in order to maintain a given performance criterion, we therefore cannot be sure to what extent the available information in central and peripheral vision was actually used or needed for the task. With object-in-scene search, \citeA{Nuthmann.VisCogn.2013,Nuthmann.JExpPsycholHuman.2014} showed that fixation durations increase both with central and peripheral low-pass filtering, suggesting that high frequencies are needed in both scene regions for efficient search. Thus, in the present experiment, object-in-scene search was given as a task as well, which implies more top-down control and predictability of what viewers do during each fixation~\cite{Henderson.PsychonBullRev.2009,Tatler.JVis.2011}. Search involves two processes that possibly occur in parallel. First, viewers need to analyze the fixated stimulus to decide whether it is the target. This process involves object identification at least to the degree that a rejection decision is possible. Second, if the fixated stimulus is not the target, viewers need to analyze the peripheral stimulus to choose a new fixation location that likely contains the search target~\cite{vanDiepen.INBOOK.1998}. The importance of different spatial frequencies for accomplishing these tasks during fixation is assessed better with a search than a memorization task.

Search studies show that viewers locate scene regions with features similar to the search target~\cite{Hwang.JVis.2009} or regions likely containing the target according to scene context~\cite{Neider.VisionRes.2006,Spotorno.JVis.2014}. Consequently, search is more efficient when the target is cued with pictures rather than words~\cite{Malcolm.JVis.2009,Nuthmann.JVis.2016}, and when targets are located at predictable compared with unpredictable locations with respect to scene context~\cite{Malcolm.JVis.2010}. Higher search efficiency in these studies was reflected in both shorter scanning times and shorter verification times. Scanning time reflects the time it takes to locate the target in the scene (i.e., fixate it for the first time), and verification time reflects the time it takes to verify the identity of the target once it has been fixated.~\citeA{Nuthmann.JVis.2016} also showed that removing color from the scene increases scanning and verification times and impairs performance by decreasing search accuracy and increasing search times.~\citeA{Castelhano.PsychonBullRev.2008} reported that target typicality only modulated verification times but not scanning times in object search arrays, with faster verification when targets were more (proto)typical for their object category. Moreover, attenuating high spatial frequencies has been shown to impede search performance differently when applied to central or peripheral vision, with peripheral low-pass filtering increasing scanning times and central low-pass filtering increasing verification times~\cite{Nuthmann.VisCogn.2013,Nuthmann.JExpPsycholHuman.2014}. These results suggest that peripheral vision aids object localization, whereas central vision aids object verification.

The present work aims to extend the findings by~\citeA{Nuthmann.VisCogn.2013,Nuthmann.JExpPsycholHuman.2014} by qualifying whether low or high spatial frequencies are more important in central and peripheral vision for object-in-scene search. To this end, we applied gaze-contingent high-pass or low-pass filters to central or peripheral vision while viewers searched for target objects that were either present or absent in the scenes. Additionally, we investigated search in both color and grayscale scenes to assess the effect of color on the importance of low or high spatial frequencies in different parts of the visual field, thus extending the findings by~\citeA{Nuthmann.JVis.2016} on the contributions of color in central and peripheral vision during scene search. 

We expected high frequencies, which are best resolved in central vision, to be critical for object identification, as they aid object--background segregation and analysis of detail. Search performance and eye-movement behavior should thus be more impaired by central low-pass filtering than central high-pass filtering. This hypothesis is challenged by the fact that high-pass filtering inherently attenuates features like luminance, color, and contrast in addition to low spatial frequencies (see Figure~\ref{fig:stimuli}); thus, to the naive observer, high-pass filtering might appear more artificial and disruptive for processing than low-pass filtering. When applying spatial-frequency filters only to central vision, however, research indicates that high-pass filtering is more beneficial for scene categorization \cite{Peyrin.BrainCogn.2003} and for eye-movement control \cite<e.g.,>{Cajar.JVis.2016} than low-pass filtering. Object verification should take substantially longer with central filtering, especially low-pass filtering, whereas peripheral filtering should not impair target verification once the target has been located. 

In contrast, we assume that peripheral vision is critical for object localization. We thus expected scanning times to prolong only slightly with central filtering, which slows down object rejection, but to prolong severely with peripheral filtering, which impairs saccade target selection. Since high frequencies can hardly be resolved in peripheral vision, search performance and eye-movement behavior should be more impaired with peripheral high-pass than low-pass filtering.

Because color is an important feature for object search and identification \cite{Hwang.Proceedings.2007}, we expected search to be more difficult in grayscale scenes than color scenes. Although color sensitivity decreases with increasing eccentricity, cones are spread throughout the retina, with their density not decreasing much beyond 10 degrees~\cite{Wells-Gray.Eye.2016}, so that color can still be used quite effectively in the periphery (\citeNP<e.g.,>{Abramov.JOptSocAmA.1991,Hansen.JVis.2009,Johnson.AmJOptomPhysiolOptics.1986}; see also \citeNP{Nuthmann.JVis.2016}). The absence of color in peripheral vision should therefore increase scanning times, and its absence in central vision should increase verification times. In grayscale scenes, the inherent attenuation of color with high-pass filtering plays no role, so high-pass filtering was expected to be less impairing than low-pass filtering compared with color scenes. On the other hand, search with low-pass filtering should be more difficult per se when color is missing as a feature. We therefore expected smaller differences between filter types in peripheral vision and larger differences in central vision when searching grayscale scenes compared with color scenes.

\section{Methods}
The search experiment was part of a large scene corpus study with $N$=200 participants, where each participant inspected 90 scenes for a memorization task in one session and 120 different scenes for an object-in-scene search task in another session (data and analyses scripts are available on Open Science Framework, DOI: 10.17605/OSF.IO/JQ56S). Session order was counterbalanced.

\subsection{Participants}
The two hundred participants (38 male, mean age: 22.6 years, range: 16 to 40 years) were students at the University of Potsdam or pupils at local schools. Participants had normal or corrected-to-normal vision and normal color discrimination. They were naive as to the purpose of the experiment and received course credit or monetary compensation for their participation. The experiment conformed to the Declaration of Helsinki. Participants gave their written informed consent prior to the experiment.

\subsection{Apparatus}
Stimuli were presented on an iiyama VisionMasterPro 514 monitor with a resolution of 1024 $\times$ 768 pixel and a refresh rate of 150 Hz. Stimuli and response collection were controlled with Matlab (The Mathworks, Natick, MA) using the Psychophysics Toolbox~\cite{Brainard.SpatVis.1997,Kleiner.Perception.2007} and the Eyelink Toolbox~\cite{Cornelissen.BehavResMeth.2002}. Viewers were seated 60 cm (23.6 inches) away from the monitor with their head stabilized by a head-chin rest. Gaze position of the dominant eye was tracked during binocular viewing with the EyeLink 1000 system (SR Research, Ontario, Canada).

\subsection{Stimuli and design}
\begin{figure}[htbp]
    \centering
    \includegraphics[width=\textwidth]{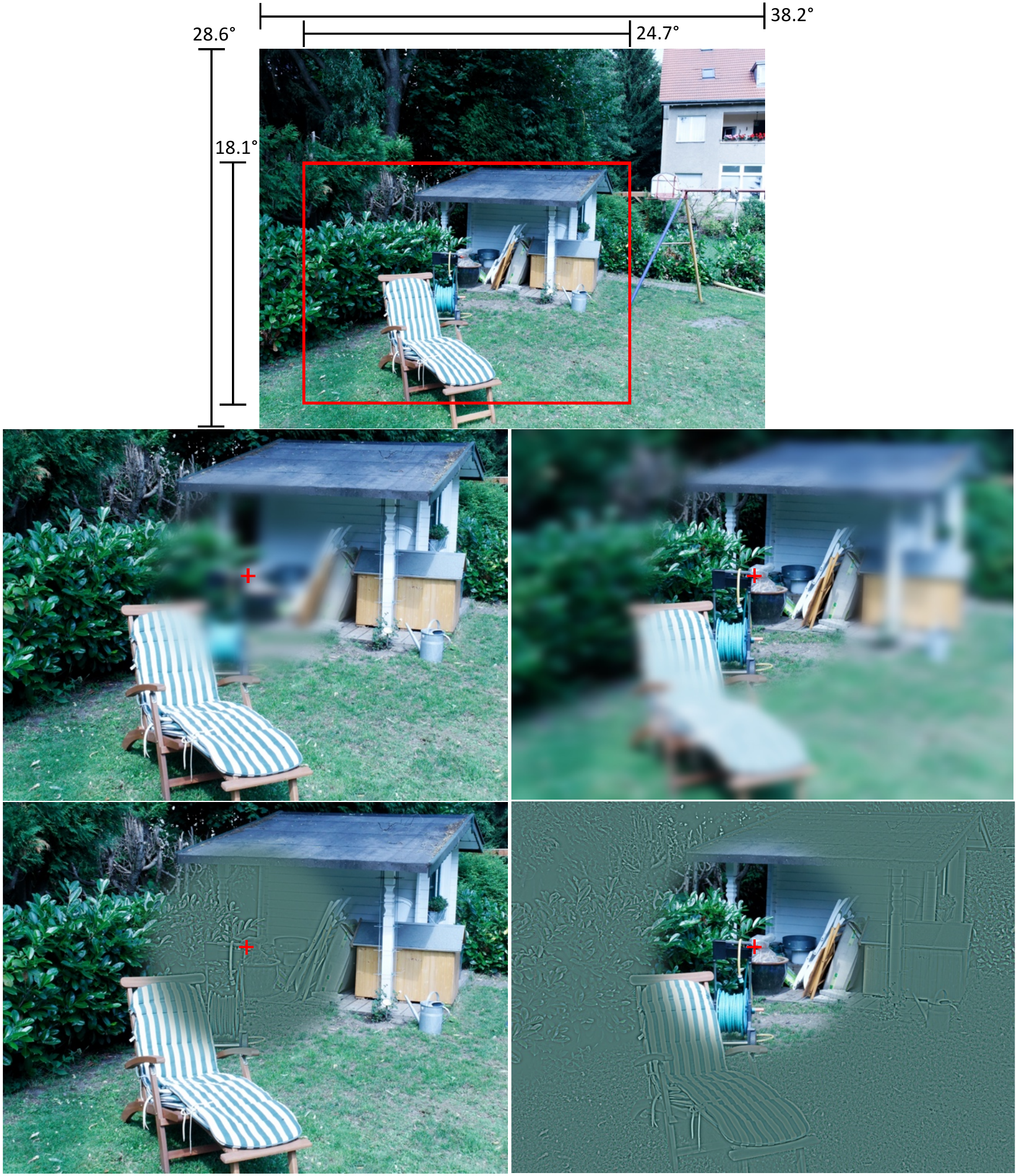}
    \caption{\label{fig:stimuli}Illustration of the five filter conditions. The red cross indicates the current gaze position. In this example, the target stimulus is the watering can (which stands on the lawn in front of the shed with the predictable location condition and on the roof of the shed with the unpredictable location condition). The top panel shows the original stimulus in the unfiltered control condition. Below, the four filter conditions are illustrated with cropped and zoomed-in versions of the original stimulus (indicated by the red frame) to better illustrate the filter effects. (\emph{Middle row, left}) Central low-pass filter. (\emph{Middle row, right}) Peripheral low-pass filter. (\emph{Bottom row, left}) Central high-pass filter. (\emph{Bottom row, right}) Peripheral high-pass filter. Unfiltered image retrieved from \url{http://info.ni.tu-berlin.de/photodb/ }\protect\cite<see> {Mohr.FrontiersPsychol.2016}.}
\end{figure}

Stimuli were 120 images of real-world scenes from the BOiS database~\cite{Mohr.FrontiersPsychol.2016} resized to 1024 $\times$ 768 pixels, subtending a visual angle of 38.2$^\circ$ $\times$ 28.6$^\circ$. Of these images, 99 depicted indoor and 21 depicted outdoor scenes. Each scene could be presented in one of three versions: with the target object present at a predictable location regarding scene context (e.g., a watering can standing on the lawn in a garden, see Figure~\ref{fig:stimuli}), an unpredictable location (e.g., a watering can standing on the roof of a shed), or absent from the scene.

For each scene, low-pass and high-pass filtered versions were prepared in advance. Filtering was realized in the Fourier domain with Gaussian filters. Cutoff frequencies for low-pass and high-pass filters were 1 c/$^\circ$ and 9 c/$^\circ$ respectively (cutoffs were defined as the half power point, where the filter response is reduced to 0.5, that is, -3 dB in the power spectrum, which corresponds to 1/$\sqrt{2}$ in the amplitude spectrum). Thus, low-pass filtering attenuated spatial frequencies above 1 c/$^\circ$ and high-pass filtering attenuated spatial frequencies below 9 c/$^\circ$. These cutoffs are near the maximal sensitivities of magno- and parvocellular cells respectively, in the lateral geniculate nucleus \cite{Derrington.JPhysiol.1984}. To give the reader a rough estimate of how far off into the periphery the cutoff frequencies were visible, we calculated the maximal eccentricities they were visible at according to the cortical magnification principle. On the superior (worst contrast sensitivity) and temporal (best constrast sensitivity) half-meridian of the visual field, the cutoff frequencies should become invisible at the eccentricities of 41.5$^\circ$ and 73.3$^\circ$ respectively for spatial frequencies of 1 c/$^\circ$, and 6.8$^\circ$ and 9.9$^\circ$ for spatial frequencies of 9 c/$^\circ$. For details about how these values were calculated, \citeA<we refer to>[p. 188]{Cajar.VisionRes.2016}.

For gaze-contingent filtering in the central or peripheral visual field, a foreground and a background image were merged in real-time using alpha blending. With central high-pass filtering, for example, the foreground image was the high-pass filtered version of the scene and the background image was the original scene. A 2D hyperbolic tangent with a slope of 0.06 served as a blending function for creating the alpha mask. The inflection point of the function corresponded to the radius of the gaze-contingent window, which was 5$^\circ$, roughly dividing central from peripheral vision \cite<see also>{Larson.JVis.2009,Loschky.JExpPsycholAppl.2002,Nuthmann.JVis.2016}. The alpha mask was centered at the current gaze position and defined the transparency value, which constitutes the weighting of the central foreground image at each point. At the point of fixation, only the foreground image was visible; with increasing eccentricity, the peripheral background image was weighted more strongly until it was fully visible.

Two filter locations (central/peripheral visual field) were crossed with two filter types (low-pass/high-pass), yielding four filter conditions: central low-pass, central high-pass, peripheral low-pass, and peripheral high-pass (for example stimuli, see Figure~\ref{fig:stimuli}). A control condition without filtering served as a baseline. This resulted in 24 trials per condition in the search task. Half of those trials were target absent and the other half target present trials; of the latter, half of the trials presented the target at a predictable and the other half at an unpredictable location. For half of the participants scenes were presented in their original color version; for the other half of the participants, the same scenes were presented in grayscale.

A Latin square design assured counterbalancing of condition--scene--target location assignments across participants. Scenes were presented in random order.

\subsection{Procedure}
At the beginning of the experiment and after every 15 trials a 9-point calibration was performed. Each trial started with a fixation check where a fixation point was presented at the center of the screen. The viewer's gaze had to stay within an area of $1.5^\circ \times 1.5^\circ$ around the fixation point for 200 ms to pass the fixation check. After a successful check, the actual trial started; if the check failed three times, a re-calibration was scheduled.

Three example trials familiarized participants with the task and the gaze-contingent window procedure. Each trial started with a picture cue of the target object on a black background presented for two seconds. After that, a black cross was presented in the center of the screen to ensure viewers always started exploring in the center of the image. After the cross was fixated, the scene was revealed. Viewers were instructed to search for the target object in the scene and to decide as fast as possible whether the object was present or absent in the scene by pressing the left or right button of the computer mouse respectively. The response deadline of 60 seconds was never actually reached in the experiment.

\subsection{Data preparation}
Saccades were detected in the raw time series of gaze positions using a velocity-based algorithm~\cite{Engbert.VisionRes.2003,Engbert.PNAS.2006} with a relative velocity threshold of 6 standard deviations and a minimum duration of 8 data samples. A total of 68 trials (0.28\%) were removed owing to poor calibration or too much data loss. Single fixations and saccades were removed if they neighbored eye blinks or were outside the monitor area. If the first or last trial event was an ongoing saccade, it was also removed. Since the present study focuses on target present trials, target absent trials were excluded from analyses. Thus, 11966 trials remained for analyses of task performance, and 115,390 fixations and 105,915 saccades remained for eye movement analyses. For analyzing fixation durations, the last ongoing fixation of each trial was excluded.

\subsection{Data analyses}
Data from trials with target objects at contextually predictable and unpredictable locations were collapsed for all analyses. For effects of target predictability on search performance and eye movements, see Figures~\ref{fig:performanceExpected}-\ref{fig:eyemovementsExpected} in the Appendix.

Separate models were run for color and grayscale scene data. Search times and eye movements were analyzed with linear mixed-effects models (LMMs) and search accuracy was analyzed with binomial generalized linear mixed-effects models (GLMMs) with a logit link function. Both LMMs and GLMMs are implemented in the lme4 package~\cite{Bates.lme4.2015}, which is supplied in the R system for statistical computing~\cite<version 3.6.0;>{RCoreTeam.2018}. Besides fixed effects for the experimental manipulations, (G)LMMs account for random effects (i.e., variance components) due to differences between participants and scenes, which reduces unexplained variance. We assumed random intercepts for participants and scenes for all models. Fixed-effects parameters were estimated via treatment contrasts, testing the effects of each of the four filter conditions against the unfiltered control condition. Selected comparisons between filter conditions were done post-hoc using paired $t$-tests. To test interactions between spatial-frequency and color filtering, we ran supplementary analyses with color as an additional fixed effect in the LMMs, providing main effects for color vs. grayscale scenes and interactions between color and spatial-frequency filtering effects. For the GLMM on search accuracies, only a main effect of color was added to the model, as a model with additional interactions yielded no stable estimates. In order to keep the presentation succinct, we only report significant effects from these supplementary analyses in the results section. 

All GLMM analyses yield regression coefficients, standard errors, $z$-values, and $p$-values for fixed effects. LMM analyses only yield regression coefficients, standard errors, and $t$-values, because the degrees of freedom are not known exactly for LMMs. For large data sets, however, the $t$-distribution has converged to the standard normal distribution for all practical purposes~\cite[Note 1]{Baayen.JMemLang.2008}. Thus, $t$-statistics exceeding an absolute value of 1.96 were considered statistically significant on the two-tailed 5\% level.

Since the distributions of all eye-movement variables and search times were positively skewed, variables were transformed before model fitting to approximate normally distributed model residuals. To find a suitable transformation, the optimal $\lambda$-coefficient for the Box-Cox power transformation~\cite{Box.JRoyalStatSoc.1964} was estimated with the \emph{boxcox} function of the \emph{MASS} package~\cite{Venables.MASS.2002}, with $y^{(\lambda)} = (y^\lambda-1)/\lambda$, if $\lambda \neq 0$ and $\log(y)$, if $\lambda = 0$. For all variables except saccade amplitudes, $\lambda$ was near zero and the log-transformation was chosen; for saccade amplitudes the exact $\lambda$ was chosen as a transformation ($\lambda$ = 0.22 for both color and grayscale scene data), yielding considerably better distributions of model residuals compared with log-transformed data.

\section{Results}

\subsection{Search performance}
\begin{figure}[htbp]
    \centering
    \includegraphics[width=0.6\textwidth]{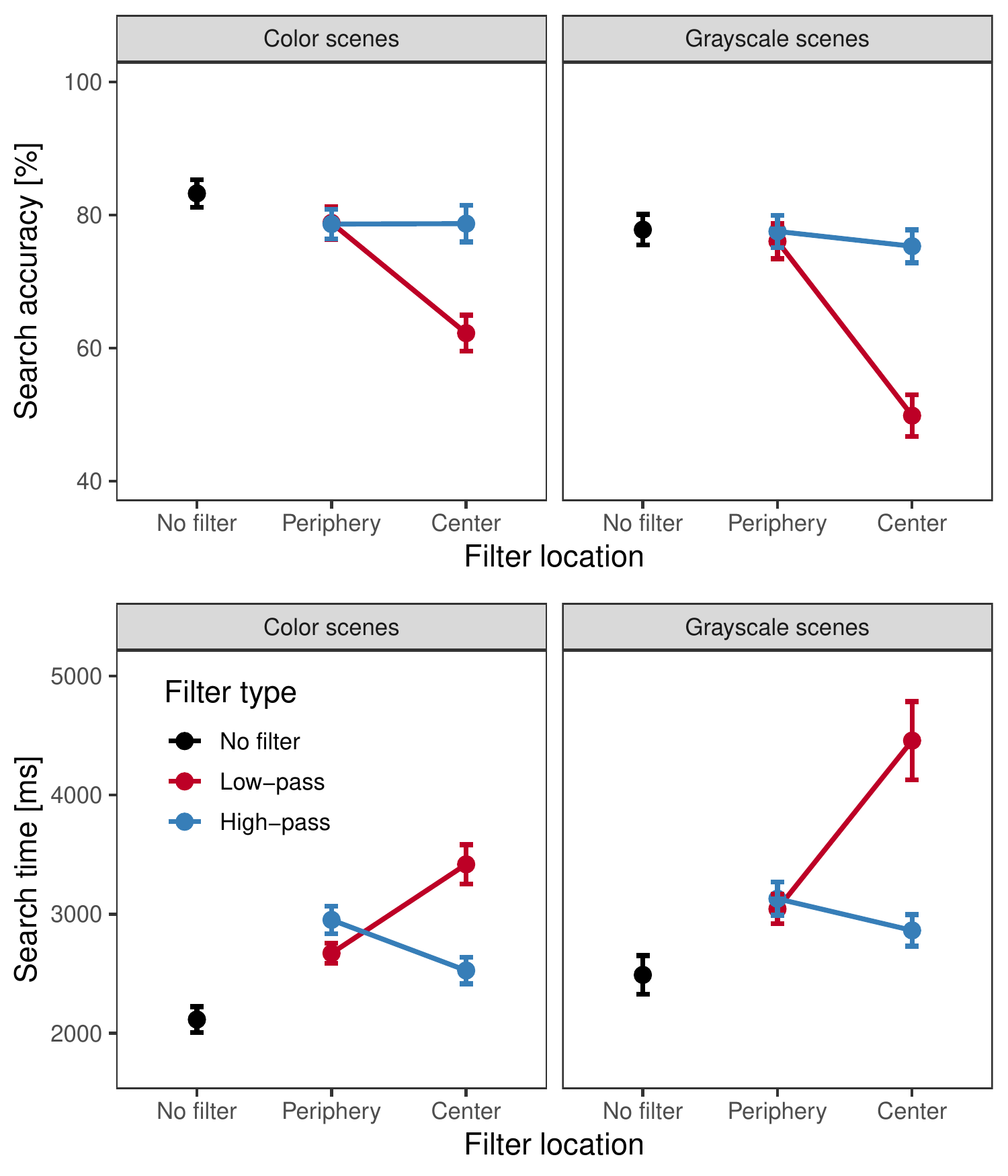}
    \caption{\label{fig:performance}Search performance for color and grayscale scenes. \emph{(Top row)} Mean search accuracies. \emph{(Bottom row)} Mean search times. Error bars represent within-subjects 95\% confidence intervals with Cousineau-Morey correction \protect \cite{Cousineau.TQMP.2005,Morey.TQMP.2008}.}
\end{figure}

In general, a similar pattern of results was observed with color and grayscale scenes regarding the effects of spatial-frequency filtering on search accuracy and search times (see Figure~\ref{fig:performance}). Performance was overall better, however, when searching color scenes than grayscale scenes, reflected in higher accuracies ($b = 0.62, SE = 0.09, t = 6.88, p < .001$) and shorter search times ($b = -0.22, SE = 0.04, t = -5.39$).

\subsubsection{Search accuracies}
In the unfiltered control condition, mean search accuracy was 83\% with color scenes and 78\% with grayscale scenes. These mean accuracies are rather low, as they reflect collapsed data from search trials with targets at predictable and unpredictable locations, and viewers performed worse with the latter (see Figure~\ref{fig:performanceExpected} in the Appendix). When searching color scenes, accuracy decreased with all filter conditions. The decrease was pronounced with central low-pass filtering ($b = -1.45, SE = 0.11, z = -12.92, p < .001$) and mild in all other conditions ($b = -0.38, SE = 0.12, z = -3.20, p = .001$ for peripheral low-pass filtering; $b = -0.38, SE = 0.12, z = -3.24, p = .001$ for peripheral high-pass filtering; $b = -0.37, SE = 0.12, z = -3.14, p = .002$ for central high-pass filtering). The decrease in accuracy with central low-pass filtering seemed even more dramatic when viewers searched grayscale scenes ($b = -1.78, SE = 0.11, z = -16.27, p < .001$), where accuracy dropped to chance performance.  With all other filter conditions in grayscale scenes, accuracy did not differ from the control condition ($b = -0.13, SE = 0.11, z = -1.17, p = .240$ for peripheral low-pass filtering; $b = -0.13, SE = 0.11, z = -0.18, p = .854$ for peripheral high-pass filtering; $b = -0.18, SE = 0.11, z = -1.56, p = .119$ for central high-pass filtering).

\subsubsection{Search times}
With all filter conditions, search times increased compared with the unfiltered control condition, showing similar patterns for color and grayscale scenes. The increase in search times was weakest with central high-pass filtering ($b = 0.16, SE = 0.02, t = 8.40$ for color scenes and $b = 0.12, SE = 0.02, t = 6.15$ for grayscale scenes) and strongest with central low-pass filtering ($b = 0.43, SE = 0.02, t = 22.18$ for color scenes and $b = 0.55, SE = 0.02, t = 28.20$ for grayscale scenes). The increase in search times with central low-pass filtering was considerably higher when searching grayscale than color scenes ($b = -0.13, SE = 0.03, t = -4.17$), complementing the stronger decrease in search accuracy with grayscale scenes. The increase of search times with peripheral filters was intermediate between the increase with central low-pass and high-pass filters ($b = 0.25, SE = 0.02, t = 13.12$ and $b = 0.34, SE = 0.02, t = 17.72$ for peripheral low-pass and high-pass filters in color scenes respectively; $b = 0.22, SE = 0.02, t = 10.96$ and $b = 0.22, SE = 0.02, t = 11.41$ for peripheral low-pass and high-pass filters in grayscale scenes respectively). The increase in search times with peripheral high-pass filtering was stronger when searching color scenes compared with grayscale scenes ($b = 0.12, SE = 0.03, t = 3.88$). Post-hoc comparisons showed a significant difference between filter types with peripheral filtering in color scenes ($t(99) = 4.04, p < .001, d = 0.09$), but not in grayscale scenes ($t(99) = 0.38, p = .702, d = 0.01$).

\subsection{Search epochs}
In order to better understand the effects of spatial-frequency filtering on sub-processes of search and thus the effects on task performance, we decomposed the search process into scanning time (time until first fixation on the target) and verification time (time from first fixation on the target until manual response). Scanning time reflects target localization, that is, the actual search process, and verification time reflects target identification plus the manual response~\cite<see also>{Castelhano.JExpPsycholHuman.2008,Malcolm.JVis.2009,Malcolm.JVis.2010,Nuthmann.VisCogn.2013,Nuthmann.JExpPsycholHuman.2014}. For classifying fixations as being on the target object or not, a polygon was drawn outlining the target and then enlarged by 1$^\circ$. Thus, fixations landing 1$^\circ$ to the edge of the object still counted as target fixations, which accounts for oculomotor errors and eye tracker inaccuracies, but ensures that near-foveal processing of the target is still possible.

\begin{figure}[htbp]
    \centering
    \includegraphics[width=\textwidth]{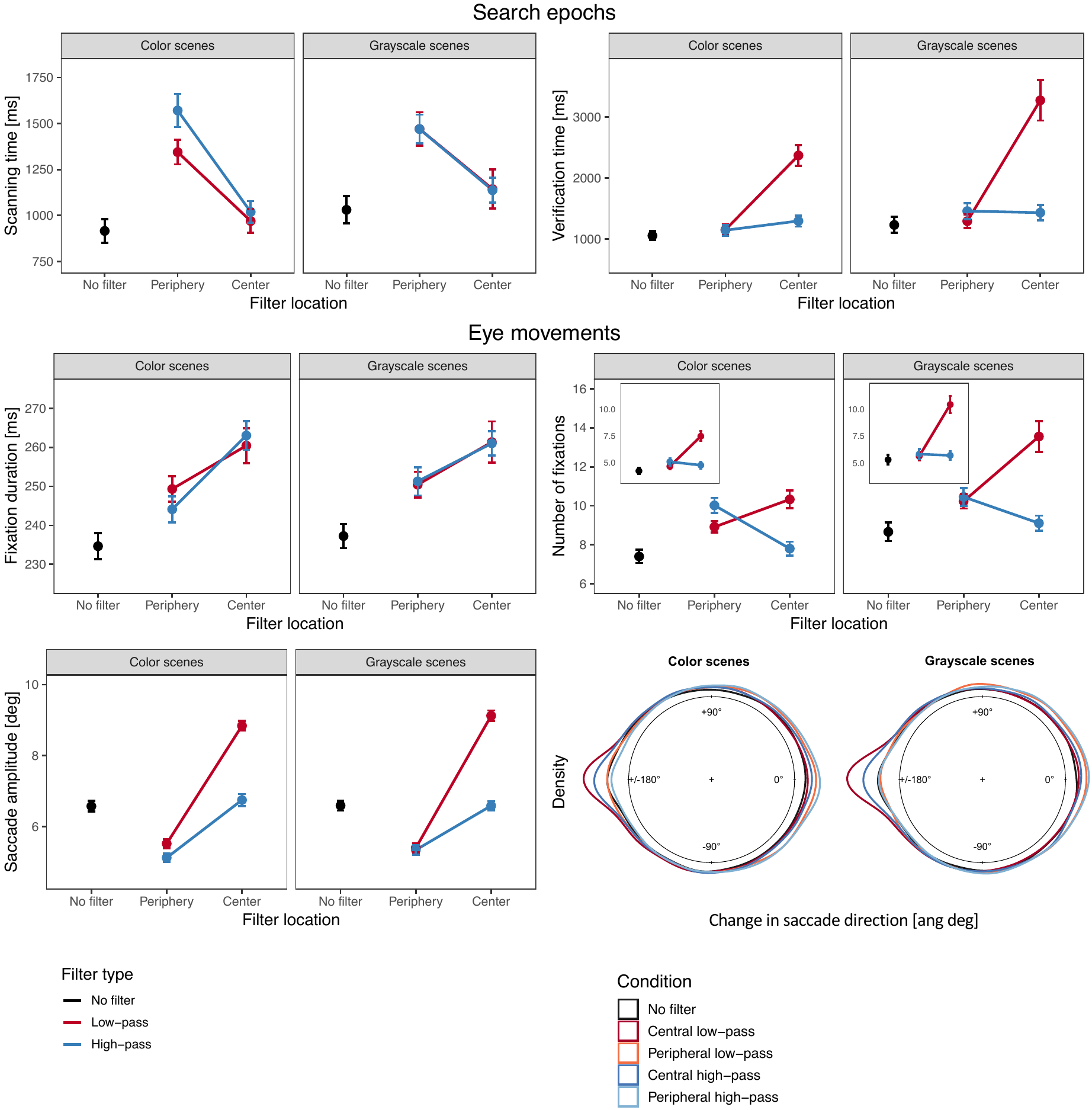}
    \caption{\label{fig:epochseye}Search epochs and eye movements with the five filter conditions for color and grayscale scenes. \emph{(Top row)} Mean scanning times \emph{(left)} and mean verification times \emph{(right)}. \emph{(Middle row)} Mean fixation durations \emph{(left)} and mean number of fixations \emph{(right)}. The inset figures with number of fixations display the mean number of fixations during verification time only. \emph{(Bottom row, left)} Mean saccade amplitudes. Error bars represent within-subjects 95\% confidence intervals with Cosineau-Morey correction \protect \cite{Cousineau.TQMP.2005,Morey.TQMP.2008}. \emph{(Bottom row, right)} Changes in saccade direction in angular degrees between successive saccades during verification time only. Changes of 0$^\circ$ and 180$^\circ$ reflect forward and backward saccades respectively; positive and negative changes reflect upward and downward saccades respectively.}
\end{figure}

As with search accuracies and search times, search epochs showed a similar pattern for color and grayscale scenes, but somewhat worse performance when searching grayscale scenes, reflected in longer scanning and verification times (see Figure~\ref{fig:epochseye}).

\subsubsection{Scanning times}
Overall, scanning times were slightly shorter when searching color scenes compared with grayscale scenes ($b = -0.08, SE = 0.04, t = -2.04$). With both color and grayscale scenes, scanning times increased considerably with peripheral filtering, but only slightly with central filtering compared with the unfiltered control condition. In color scenes, scanning times did not increase with central low-pass filtering ($b = 0.02, SE = 0.03, t = 0.77$) and only slightly with central high-pass filtering ($b = 0.06, SE = 0.03, t = 2.10$). With peripheral filtering, however, scanning times increased strongly ($b = 0.43, SE = 0.03, t = 15.41$ for peripheral low-pass filtering; $b = 0.56, SE = 0.03, t = 20.15$ for peripheral high-pass filtering), with significantly longer scanning times for peripheral high-pass filtering than low-pass filtering ($t(99) = 3.61, p < .001, d = 0.13$). Similar effects emerged with searching grayscale scenes. Scanning times increased slightly with both central low-pass and high-pass filtering ($b = 0.06, SE = 0.03, t = 2.13$ and $b = 0.08, SE = 0.03, t = 2.82$ respectively), but increased more strongly with peripheral filtering ($b = 0.40, SE = 0.03, t = 13.67$ for peripheral low-pass filtering; $b = 0.35, SE = 0.03, t = 12.09$ for peripheral high-pass filtering). The increase in scanning times with peripheral high-pass filtering was stronger in color scenes than grayscale scenes ($b = 0.20, SE = 0.05, t = 4.36$). Contrary to searching color scenes, there was no effect of filter type on scanning times when searching grayscale scenes, neither with central nor peripheral filtering (see Figure~\ref{fig:epochseye}, top left graph).

\subsubsection{Verification times}
Verification times were rather long in the present experiment, because scenes were complex, targets were located at unpredictable locations in half of the trials, and verification times included the manual response to the target.

Verification times increased in most filter conditions compared with search in unfiltered scenes. Overall, verification times were shorter when searching color compared with grayscale scenes ($b = -0.20, SE = 0.05, t = -3.71$). For search in color scenes, verification times increased slightly with peripheral filtering ($b = 0.06, SE = 0.03, t = 2.08$ for peripheral low-pass filtering; $b = 0.07, SE = 0.03, t = 2.34$ for peripheral high-pass filtering). Verification times increased more strongly with central high-pass filtering ($b = 0.19, SE = 0.03, t = 6.62$) and drastically with central low-pass filtering ($b = 0.69, SE = 0.03, t = 24.12$). This pattern was similar when searching grayscale scenes, with no effect of peripheral low-pass filtering on verification times ($b = 0.03, SE = 0.03, t = 1.11$), and a moderate increase of verification times with high-pass filtering in central and peripheral vision ($b = 0.11, SE = 0.03, t = 3.94$ and $b = 0.09, SE = 0.03, t = 3.28$ respectively). The strong increase of verification times with central low-pass filtering compared with the control condition ($b = 0.86, SE = 0.03, t = 28.97$) was even higher when searching grayscale scenes than color scenes ($b = -0.17, SE = 0.04, t = -4.06$). Post-hoc comparisons showed no difference in verification times between peripheral filter types with grayscale scenes ($t(99) = 1.88, p = .006, d = 0.06$).

\subsection{Eye movements}
As with task performance and search epochs, effects of filtering on eye-movement behavior were similar with search in color and grayscale scenes, with stronger modulations of behavior when searching grayscale scenes (see Figure~\ref{fig:epochseye}).

\subsubsection{Fixation durations}
In both color and grayscale scenes, mean fixation durations increased with all filter conditions compared with the unfiltered control condition. Central filtering led to a stronger increase of fixation durations than peripheral filtering. For search in color scenes, central low-pass and high-pass filtering prolonged fixation durations similarly ($b = 0.09, SE = 0.01, t = 13.34$ and $b = 0.10, SE = 0.01, t = 14.50$ respectively). Numerically, the increase was stronger with high-pass than low-pass filtering, but the difference between central filter types was only marginally significant ($t(99) = 1.79, p = .076, d = 0.02$). The increase of fixation durations with peripheral filtering compared to the control condition was stronger with peripheral low-pass filtering ($b = 0.06, SE = 0.01, t = 9.17$) than with peripheral high-pass filtering ($b = 0.04, SE = 0.01, t = 6.67$). This difference between peripheral filter types was statistically significant ($t(99) = 2.51, p = .014, d = 0.02$). When searching grayscale scenes, there was no effect of filter type on fixation durations (see Figure~\ref{fig:epochseye}), which increased similarly with central low-pass and high-pass filtering ($b = 0.07, SE = 0.01, t = 12.70$ and $b = 0.08, SE = 0.01, t = 13.20$ respectively) and with peripheral low-pass and high-pass filtering ($b = 0.06, SE = 0.01, t = 8.93$ and $b = 0.06, SE = 0.01, t = 9.18$ respectively). There was no main effect of color and there were no interaction effects between spatial-frequency filtering and color on fixation durations.

\subsubsection{Number of fixations}
Viewers needed fewer fixations to find the target in color than in grayscale scenes ($b = -0.22, SE = 0.03, t = -6.23$). With color scenes, mean number of fixations was highest with central low-pass filtering ($b = 0.28, SE = 0.02, t = 13.72$) and peripheral high-pass filtering ($b = 0.30, SE = 0.02, t = 14.89$), with no significant difference between the two ($t(99) = 0.73, p = .465, d = 0.02$). Peripheral low-pass filtering also increased the number of fixations considerably ($b = 0.20, SE = 0.02, t = 9.90$), whereas central high-pass filtering increased number of fixations only slightly ($b = 0.04, SE = 0.02, t = 2.06$) compared with the unfiltered control condition. With grayscale scenes, a similar pattern emerged. Central high-pass filtering increased the number of fixations only slightly ($b = 0.04, SE = 0.02, t = 2.01$), whereas central low-pass filtering increased the number of fixations strongly ($b = 0.43, SE = 0.02, t = 21.50$), the increase being a lot higher than with color scenes ($b = -0.16, SE = 0.03, t = -5.03$). Also contrary to searching color scenes, the increase in the number of fixations with peripheral filters was similar with low-pass and high-pass filters ($b = 0.18, SE = 0.02, t = 8.90$ and $b = 0.17, SE = 0.02, t = 8.67$ respectively). Peripheral high-pass filtering did not increase the number of fixations as strongly compared with the control condition as it did in color scenes ($b = 0.12, SE = 0.03, t = 3.99$).

The inset figures for mean number of fixations (see Figure \ref{fig:epochseye}) show the mean number of fixations during verification time. It appears that viewers make considerably more fixations during object verification with central low-pass filtering than with any of the other filter conditions, especially when searching grayscale scenes.

\subsubsection{Saccade amplitudes}
The modulation of saccade amplitudes by spatial-frequency filtering was similar with color and grayscale scenes, although overall, amplitudes were shorter when searching color scenes ($-0.02, SE = 0.01, t = -3.96$). Saccade amplitudes were not affected by central high-pass filtering ($b = 0.01, SE = 0.004, t = 1.94$ for color scenes and $b = -0.001, SE = 0.004, t = -0.26$ for grayscale scenes), but strongly increased with central low-pass filtering ($b = 0.14, SE = 0.004, t = 36.55$ for color scenes and $b = 0.16, SE = 0.003, t = 47.29$ for grayscale scenes). Peripheral filtering, on the other hand, shortened saccade amplitudes compared with the control condition: For searching color scenes, the decrease in saccade amplitudes was slightly stronger with high-pass ($b = -0.06, SE = 0.004, t = -14.78$) than with low-pass filtering ($b = -0.05, SE = 0.004, t = -11.63$). This difference between peripheral filter types was statistically significant ($t(99) = 2.50, p = .014, d = 0.01$). For searching grayscale scenes, saccade amplitudes also decreased ($b = -0.05, SE = 0.004, t = -15.01$ for low-pass filtering and $b = -0.05, SE = 0.004, t = -13.23$ for high-pass filtering), but the difference between peripheral filter types was not significant ($t(99) = -1.59, p = .116, d = -0.01$).

\subsubsection{Changes in saccade direction}
We expected target verification to be more difficult when central vision was impaired by spatial-frequency filters (especially low-pass filters) than when central vision was left intact. Viewers might therefore adopt different strategies for identifying the potential target object in the different filter conditions. To reveal viewers' exploration strategies during target verification, we inspected changes in saccade direction in the different filter conditions, that is, saccadic angles between two successive saccades. The bottom right graph in Figure \ref{fig:epochseye} shows the distributions of changes in saccade direction for the different filter conditions during target verification. Angles of 0$^\circ$ reflect forward saccades and angles of 180$^\circ$ reflect backward or return saccades; positive angles reflect upward and negative angles reflect downward saccades. Distributions show that central low-pass filtering involved a considerably higher amount of return saccades during target verification than the other filter conditions and the control condition, especially with search in grayscale scenes. To a weaker extent, this effect also emerged with central high-pass filtering. 

\section{Discussion}
In the present study we investigated the importance of spatial frequencies and color in central and peripheral vision for eye-movement control during scene search. Results show considerable differences in both search performance and eye-movement behavior between high and low spatial-frequency filters, especially with search in color scenes and filters in central vision. Overall, performance and eye-movement behavior were impaired more strongly when searching grayscale scenes than color scenes~\cite<see also>{Nuthmann.JVis.2016}. Interestingly, color information was particularly important in the low-frequency band in peripheral vision.

\subsubsection{Peripheral filtering} 
Compared with searching unfiltered scenes, peripheral spatial-frequency filtering slightly decreased search accuracies with color scenes, but not at all with grayscale scenes. Search times, however, increased considerably with both color and grayscale scenes. This effect mainly derives from an increase in scanning times, indicating that it took longer to locate the target object in the scene. Object verification time, once the target object was located, was hardly affected by peripheral filtering. This finding is not surprising given that central vision is critical for object identification~\cite{Henderson.PerceptPsychophys.2003} and information in the central visual field was unaltered. The increase in scanning times was caused by a combination of shorter saccade amplitudes and more as well as longer fixations. The decrease in saccade amplitudes replicates previous findings~ \cite{Cajar.VisionRes.2016,Cajar.JVis.2016,Foulsham.AttenPerceptPsychophys.2011,Laubrock.JVis.2013,Loschky.JExpPsycholAppl.2002,Loschky.VisCogn.2005,Nuthmann.VisCogn.2013} and likely reflects a shrinkage of the attentional focus to the unfiltered central region~\cite{Cajar.JVis.2016}. This tunnel-vision effect also accounts for the higher number of fixations~\cite<see also>{Foulsham.AttenPerceptPsychophys.2011,Nuthmann.VisCogn.2013,Nuthmann.JExpPsycholHuman.2014}, as the scene is scanned for the target in smaller steps. Fixation durations increased because saccade target selection took longer with higher peripheral processing difficulty. In color scenes, as expected, search performance and eye-movement behavior were affected more severely by high-pass than by low-pass filtering, reflected in longer search and scanning times, a higher number of fixations and slightly shorter saccade amplitudes. Fixation durations were longer with low-pass filtering, suggesting that more time was invested for selecting a new peripheral saccade target when the available information was easier to process~\cite<see also>{Cajar.VisionRes.2016,Cajar.JVis.2016,Laubrock.JVis.2013}.

Surprisingly the benefit for low spatial frequencies in peripheral vision disappeared completely in all search and eye-movement parameters when viewers searched grayscale scenes. This result suggests that the benefit for object search with peripheral low-pass filtering mainly arose from the availability of coarse-grained color information (which is strongly attenuated with high-pass filtering) rather than the availability of coarse-grained luminance information. This result is compatible with \citeA{Hwang.Proceedings.2007}, who found that color dominates visual guidance when searching for scene patches and reduces the guidance by other stimulus dimensions such as intensity or contrast. A strong influence of color in peripheral vision might be specific to scene search, though---in a previous study with scene memorization and peripheral object detection in grayscale scenes, we found considerable differences both in task performance and eye-movement behavior between peripheral low-pass and high-pass filters \cite{Cajar.JVis.2016}.

\subsubsection{Central filtering}
With spatial-frequency filtering in central vision, effects were rather different. First, object localization was hardly affected, whereas effects on object identification were pronounced \cite<see also>{Nuthmann.JExpPsycholHuman.2014}. Second, search performance and eye-movement behavior depended strongly on filter type in both color and grayscale scenes. With central high-pass filtering, search accuracies were not affected at all in grayscale scenes and only slightly decreased in color scenes compared with the unfiltered control. Search times did increase with high-pass filtering, but with the least increase of the four filter conditions. The longer search times derive from a combination of slight to moderate increases of scanning and verification times. Saccade amplitudes were not affected at all by high-pass filtering, and the number of fixations hardly increased. Higher processing difficulty with high-pass filtering was mainly counteracted by increasing fixation durations, thus taking more time to process the fixated stimulus. Overall, performance and eye-movement behavior with central high-pass filtering were nearest-to-normal. Although high-pass filters inherently attenuate luminance, contrast, and color in the scene, they affected object localization and verification only slightly when applied to central vision, which is an interesting new finding of the present study.

Central low-pass filtering, on the other hand, turned out to be the most detrimental filter condition. Search accuracies decreased drastically and even dropped to chance performance when searching grayscale scenes. Search times also increased more than in all other conditions. It appears that this decrease in search performance originates from the difficulty of identifying the target object once it had been found: scanning times increased only slightly, as with central high-pass filtering, but target verification times increased drastically compared with the control and all other filter conditions. The effect was larger when searching grayscale scenes. Eye-movement behavior was also modulated strongly by central low-pass filtering, with increased fixation durations, number of fixations, and saccade amplitudes. The increase in saccade amplitudes replicates previous findings~\cite{Cajar.VisionRes.2016,Cajar.JVis.2016,Laubrock.JVis.2013,Nuthmann.JExpPsycholHuman.2014}, suggesting that viewers show an attentional bias toward the visual periphery with a low-pass filtered center \cite{Cajar.JVis.2016}. The number of fixations increased with central low-pass filtering mainly because more fixations were made during target verification than in all other conditions, especially with grayscale scenes. Furthermore, a higher amount of return saccades during target verification was observed compared with the other conditions, as was reported by~\citeA{Henderson.PerceptPsychophys.1997} with a central scotoma in an array of line drawings of objects. These findings likely reflect a strategy of saccading back and forth between the target object in the strongly blurred center and neighboring scene regions to try and confirm object identity with the help of the less blurred periphery. Almost all effects were stronger when searching grayscale scenes, where search performance dropped down to chance level. Thus, object identification without high and medium spatial frequencies was extremely challenging, and nearly impossible when color was unavailable as a diagnostic object feature.

In contrast with peripheral filtering, where effects of filter type completely disappeared with grayscale scenes, differences between central filter types were even larger with grayscale than color scenes. Thus, the specific spatial-frequency content in central vision became even more important when color was removed from the scene. This new result of the present study suggests that the proposed dominance of color over other stimulus dimensions during search \cite{Hwang.Proceedings.2007} might hold for the process of object localization in peripheral vision, but not for object identification in central vision. 

\subsection{Conclusions}
The present work demonstrates that object identification in central vision mainly depends on high spatial frequencies and is nearly impossible when based solely on low frequencies. For object localization in peripheral vision the type of available spatial-frequency information seems less important, whereas color information is critical for facilitating search. We corroborate previous findings that central vision is critical for object identification and peripheral vision is critical for object localization. Our results highlight the different roles of central and peripheral vision to object-in-scene search and provide a basis for characterizing the contributions of different spatial-frequency bands for search. 
\newpage

\section{Acknowledgments}
This work was funded by Deutsche Forschungsgemeinschaft (grants LA 2884/1 to J. L. and EN 471/10 to R. E.). We thank Petra Schienmann and our student assistants for their help during data collection and Lisa Buchwald for object annotations in the scenes.

\section{Appendix}
\begin{figure}[htbp]
   \centering
   \includegraphics[width=\textwidth]{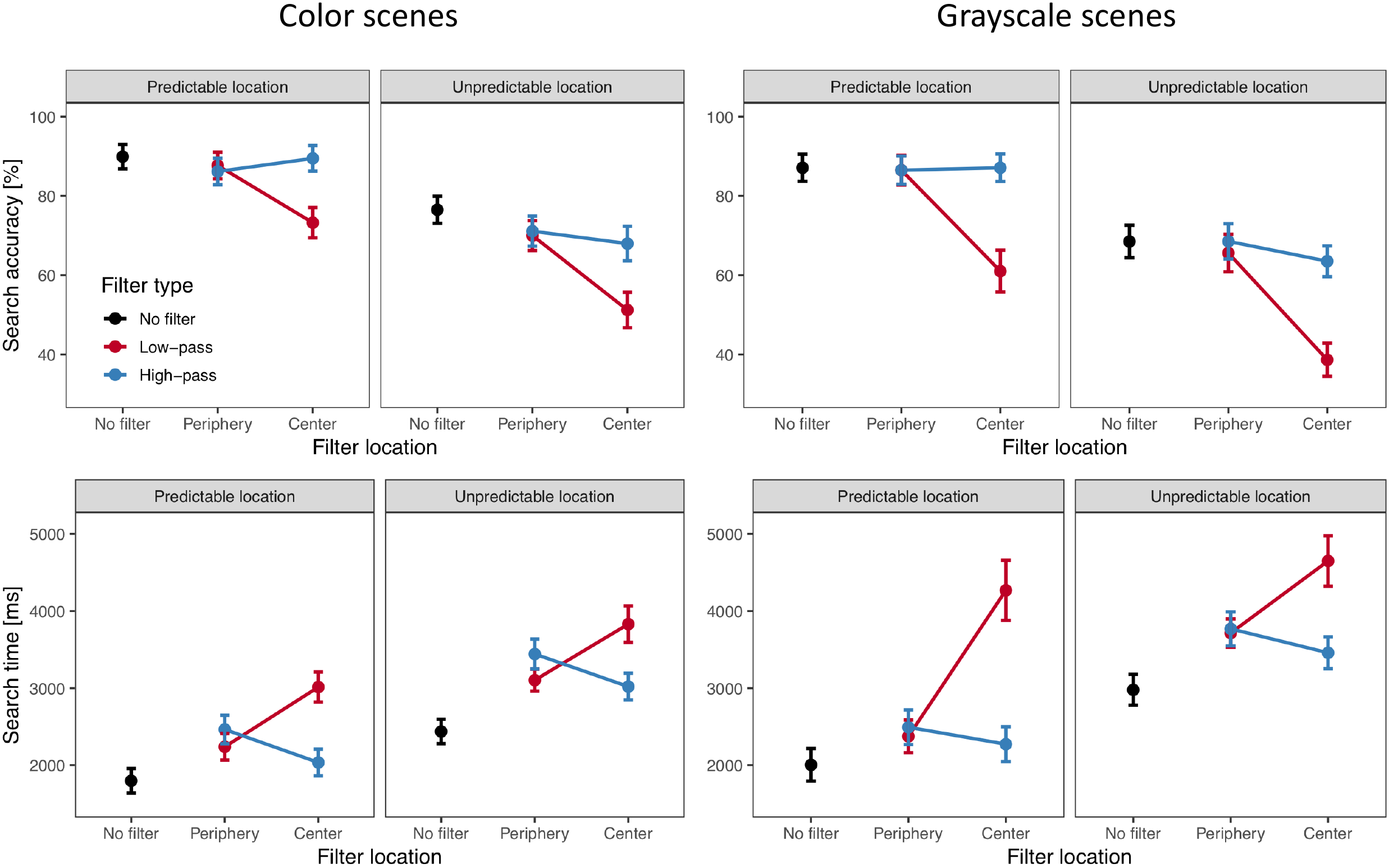}
   \caption{\label{fig:performanceExpected}Search accuracies \emph{(upper row)} and search times \emph{(lower row)} for the five filter conditions in color scenes \emph{(left)} and grayscale scenes \emph{(right)}, with targets at contextually predictable or unpredictable locations. Error bars represent within-subjects 95\% confidence intervals with Cousineau-Morey correction \protect \cite{Cousineau.TQMP.2005,Morey.TQMP.2008}.}
\end{figure}

\begin{figure}[htbp]
   \centering
   \includegraphics[width=\textwidth]{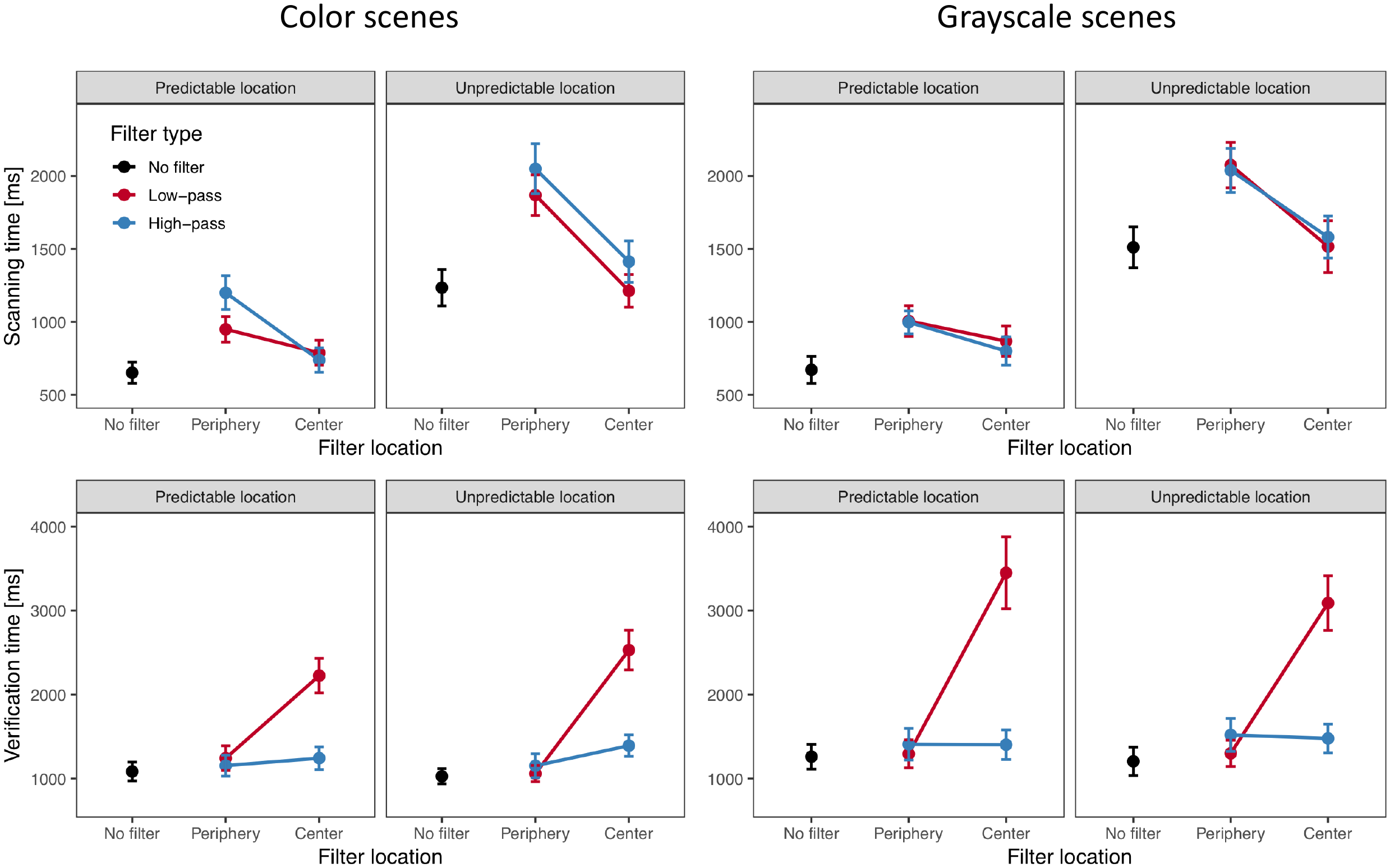}
   \caption{\label{fig:epochsExpected}Scanning times \emph{(upper row)} and verification times \emph{(lower row)} for the five filter conditions in color scenes \emph{(left)} and grayscale scenes \emph{(right)}, with targets at contextually predictable or unpredictable locations. Error bars represent within-subjects 95\% confidence intervals with Cousineau-Morey correction \protect \cite{Cousineau.TQMP.2005,Morey.TQMP.2008}.}
\end{figure}

\begin{figure}[ht!]
   \centering
   \includegraphics[width=\textwidth]{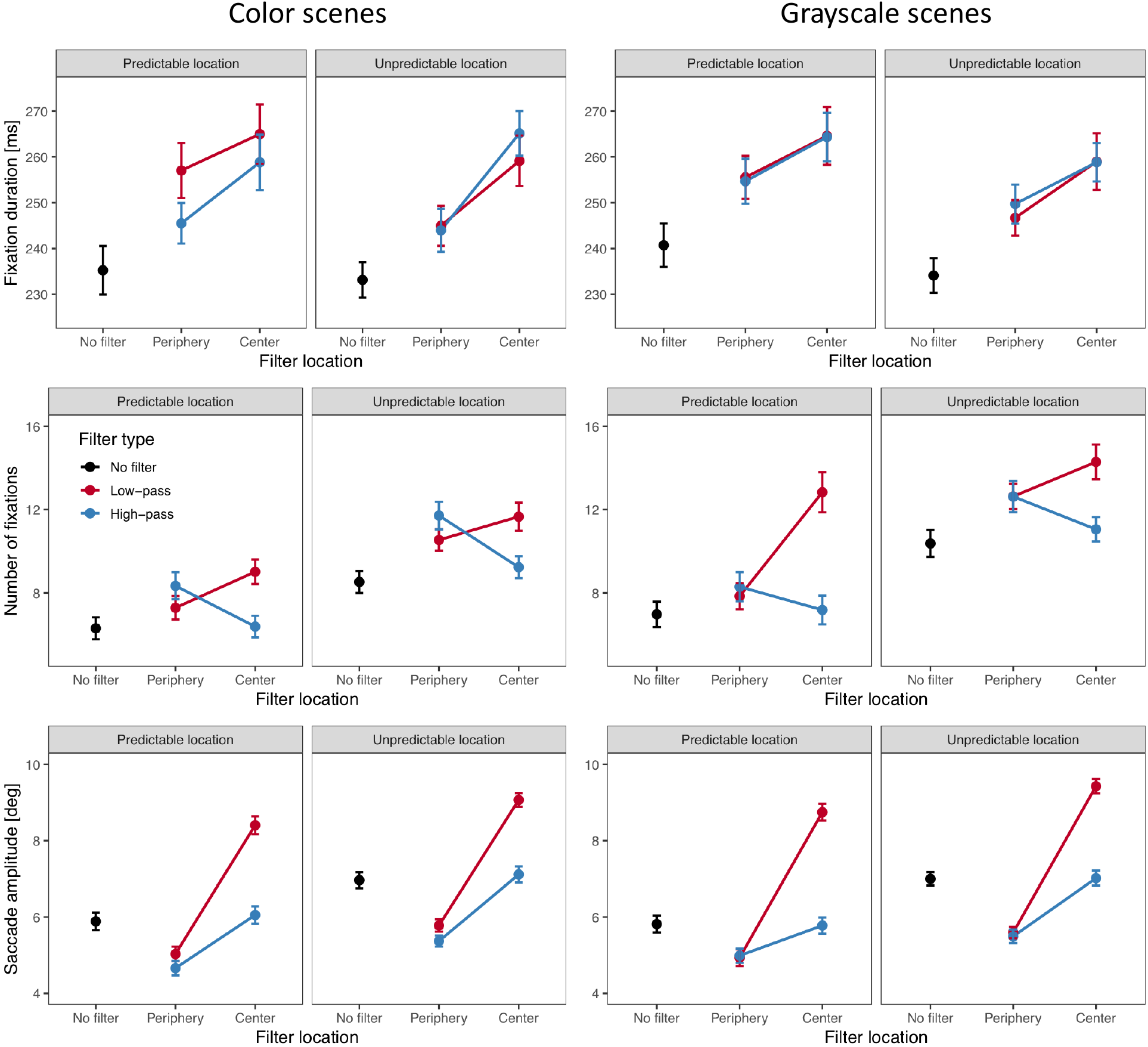}
   \caption{\label{fig:eyemovementsExpected}Fixation durations \emph{(top row)}, number of fixations \emph{(middle row)}, and saccade amplitudes \emph{(bottom row)} in the five filter conditions for color scenes \emph{(left)} and grayscale scenes \emph{(right)}, with targets at contextually predictable or unpredictable locations. Error bars represent within-subjects 95\% confidence intervals with Cousineau-Morey correction \protect \cite{Cousineau.TQMP.2005,Morey.TQMP.2008}.}
\end{figure}

The following figures show the effects of target predictability (i.e., target location is either contextually predictable or unpredictable) on search performance (Figure \ref{fig:performanceExpected}), search epochs (Figure \ref{fig:epochsExpected}), and eye movements (Figure \ref{fig:eyemovementsExpected}).

\vspace{1cm}

\clearpage

\bibliography{references}

\begin{thebibliography}{}

\bibitem [\protect \citeauthoryear {%
Abramov%
, Gordon%
\BCBL {}\ \BBA {} Chan%
}{%
Abramov%
\ \protect \BOthers {.}}{%
{\protect \APACyear {1991}}%
}]{%
Abramov.JOptSocAmA.1991}
\APACinsertmetastar {%
Abramov.JOptSocAmA.1991}%
\begin{APACrefauthors}%
Abramov, I.%
, Gordon, J.%
\BCBL {}\ \BBA {} Chan, H.%
\end{APACrefauthors}%
\unskip\
\newblock
\APACrefYearMonthDay{1991}{}{}.
\newblock
{\BBOQ}\APACrefatitle {Color appearance in the peripheral retina: effects of
  stimulus size} {Color appearance in the peripheral retina: effects of
  stimulus size}.{\BBCQ}
\newblock
\APACjournalVolNumPages{Journal of the Optical Society of America
  A}{8}{2}{404--414}.
\newblock
\begin{APACrefDOI} \doi{10.1364/JOSAA.8.000404} \end{APACrefDOI}
\PrintBackRefs{\CurrentBib}

\bibitem [\protect \citeauthoryear {%
Baayen%
, Davidson%
\BCBL {}\ \BBA {} Bates%
}{%
Baayen%
\ \protect \BOthers {.}}{%
{\protect \APACyear {2008}}%
}]{%
Baayen.JMemLang.2008}
\APACinsertmetastar {%
Baayen.JMemLang.2008}%
\begin{APACrefauthors}%
Baayen, R\BPBI H.%
, Davidson, D\BPBI J.%
\BCBL {}\ \BBA {} Bates, D\BPBI M.%
\end{APACrefauthors}%
\unskip\
\newblock
\APACrefYearMonthDay{2008}{}{}.
\newblock
{\BBOQ}\APACrefatitle {Mixed-effects modeling with crossed random effects for
  subjects and items} {Mixed-effects modeling with crossed random effects for
  subjects and items}.{\BBCQ}
\newblock
\APACjournalVolNumPages{Journal of Memory and Language}{59}{}{390--412}.
\newblock
\begin{APACrefDOI} \doi{10.1016/j.jml.2007.12.005} \end{APACrefDOI}
\PrintBackRefs{\CurrentBib}

\bibitem [\protect \citeauthoryear {%
Bates%
, Maechler%
, Bolker%
\BCBL {}\ \BBA {} Walker%
}{%
Bates%
\ \protect \BOthers {.}}{%
{\protect \APACyear {2015}}%
}]{%
Bates.lme4.2015}
\APACinsertmetastar {%
Bates.lme4.2015}%
\begin{APACrefauthors}%
Bates, D.%
, Maechler, M.%
, Bolker, B.%
\BCBL {}\ \BBA {} Walker, S.%
\end{APACrefauthors}%
\unskip\
\newblock
\APACrefYearMonthDay{2015}{}{}.
\newblock
{\BBOQ}\APACrefatitle {Fitting linear mixed-effects models using lme4} {Fitting
  linear mixed-effects models using lme4}.{\BBCQ}
\newblock
\APACjournalVolNumPages{Journal of Statistical Software}{67}{1}{1--48}.
\newblock
\begin{APACrefDOI} \doi{10.18637/jss.v067.i01} \end{APACrefDOI}
\PrintBackRefs{\CurrentBib}

\bibitem [\protect \citeauthoryear {%
Box%
\ \BBA {} Cox%
}{%
Box%
\ \BBA {} Cox%
}{%
{\protect \APACyear {1964}}%
}]{%
Box.JRoyalStatSoc.1964}
\APACinsertmetastar {%
Box.JRoyalStatSoc.1964}%
\begin{APACrefauthors}%
Box, G\BPBI E\BPBI P.%
\BCBT {}\ \BBA {} Cox, D\BPBI R.%
\end{APACrefauthors}%
\unskip\
\newblock
\APACrefYearMonthDay{1964}{}{}.
\newblock
{\BBOQ}\APACrefatitle {An analysis of transformations} {An analysis of
  transformations}.{\BBCQ}
\newblock
\APACjournalVolNumPages{Journal of the Royal Statistical
  Society}{26B}{}{211--252}.
\newblock
\begin{APACrefDOI} \doi{10.1111/j.2517-6161.1964.tb00553.x} \end{APACrefDOI}
\PrintBackRefs{\CurrentBib}

\bibitem [\protect \citeauthoryear {%
Brainard%
}{%
Brainard%
}{%
{\protect \APACyear {1997}}%
}]{%
Brainard.SpatVis.1997}
\APACinsertmetastar {%
Brainard.SpatVis.1997}%
\begin{APACrefauthors}%
Brainard, D\BPBI H.%
\end{APACrefauthors}%
\unskip\
\newblock
\APACrefYearMonthDay{1997}{}{}.
\newblock
{\BBOQ}\APACrefatitle {The {P}sychophysics {T}oolbox} {The {P}sychophysics
  {T}oolbox}.{\BBCQ}
\newblock
\APACjournalVolNumPages{Spatial Vision}{10}{}{433--436}.
\newblock
\begin{APACrefDOI} \doi{10.1163/156856897X00357} \end{APACrefDOI}
\PrintBackRefs{\CurrentBib}

\bibitem [\protect \citeauthoryear {%
Cajar%
, Engbert%
\BCBL {}\ \BBA {} Laubrock%
}{%
Cajar%
, Engbert%
\BCBL {}\ \BBA {} Laubrock%
}{%
{\protect \APACyear {2016}}%
}]{%
Cajar.VisionRes.2016}
\APACinsertmetastar {%
Cajar.VisionRes.2016}%
\begin{APACrefauthors}%
Cajar, A.%
, Engbert, R.%
\BCBL {}\ \BBA {} Laubrock, J.%
\end{APACrefauthors}%
\unskip\
\newblock
\APACrefYearMonthDay{2016}{}{}.
\newblock
{\BBOQ}\APACrefatitle {Spatial frequency processing in the central and
  peripheral visual field during scene viewing.} {Spatial frequency processing
  in the central and peripheral visual field during scene viewing.}{\BBCQ}
\newblock
\APACjournalVolNumPages{Vision Research}{127}{}{186--197}.
\newblock
\begin{APACrefDOI} \doi{10.1016/j.visres.2016.05.008} \end{APACrefDOI}
\PrintBackRefs{\CurrentBib}

\bibitem [\protect \citeauthoryear {%
Cajar%
, Schneewei\ss%
, Engbert%
\BCBL {}\ \BBA {} Laubrock%
}{%
Cajar%
, Schneewei\ss%
\BCBL {}\ \protect \BOthers {.}}{%
{\protect \APACyear {2016}}%
}]{%
Cajar.JVis.2016}
\APACinsertmetastar {%
Cajar.JVis.2016}%
\begin{APACrefauthors}%
Cajar, A.%
, Schneewei\ss, P.%
, Engbert, R.%
\BCBL {}\ \BBA {} Laubrock, J.%
\end{APACrefauthors}%
\unskip\
\newblock
\APACrefYearMonthDay{2016}{}{}.
\newblock
{\BBOQ}\APACrefatitle {Coupling of attention and saccades when viewing scenes
  with central and peripheral degradation} {Coupling of attention and saccades
  when viewing scenes with central and peripheral degradation}.{\BBCQ}
\newblock
\APACjournalVolNumPages{Journal of Vision}{16(2):8}{}{1--19}.
\newblock
\begin{APACrefDOI} \doi{10.1167/16.2.8} \end{APACrefDOI}
\PrintBackRefs{\CurrentBib}

\bibitem [\protect \citeauthoryear {%
Castelhano%
\ \BBA {} Henderson%
}{%
Castelhano%
\ \BBA {} Henderson%
}{%
{\protect \APACyear {2008}}%
}]{%
Castelhano.JExpPsycholHuman.2008}
\APACinsertmetastar {%
Castelhano.JExpPsycholHuman.2008}%
\begin{APACrefauthors}%
Castelhano, M\BPBI S.%
\BCBT {}\ \BBA {} Henderson, J\BPBI M.%
\end{APACrefauthors}%
\unskip\
\newblock
\APACrefYearMonthDay{2008}{}{}.
\newblock
{\BBOQ}\APACrefatitle {The influence of color on the perception of scene gist}
  {The influence of color on the perception of scene gist}.{\BBCQ}
\newblock
\APACjournalVolNumPages{Journal of Experimental Psychology: Human Perception
  and Performance}{34}{}{660--675}.
\newblock
\begin{APACrefDOI} \doi{10.1037/0096-1523.34.3.660} \end{APACrefDOI}
\PrintBackRefs{\CurrentBib}

\bibitem [\protect \citeauthoryear {%
Castelhano%
, Pollatsek%
\BCBL {}\ \BBA {} Cave%
}{%
Castelhano%
\ \protect \BOthers {.}}{%
{\protect \APACyear {2008}}%
}]{%
Castelhano.PsychonBullRev.2008}
\APACinsertmetastar {%
Castelhano.PsychonBullRev.2008}%
\begin{APACrefauthors}%
Castelhano, M\BPBI S.%
, Pollatsek, A.%
\BCBL {}\ \BBA {} Cave, K\BPBI R.%
\end{APACrefauthors}%
\unskip\
\newblock
\APACrefYearMonthDay{2008}{}{}.
\newblock
{\BBOQ}\APACrefatitle {Typicality aids search for an unspecified target, but
  only in identification and not in attentional guidance} {Typicality aids
  search for an unspecified target, but only in identification and not in
  attentional guidance}.{\BBCQ}
\newblock
\APACjournalVolNumPages{Psychonomic Bulletin \& Review}{15}{4}{795--801}.
\newblock
\begin{APACrefDOI} \doi{10.3758/PBR.15.4.795} \end{APACrefDOI}
\PrintBackRefs{\CurrentBib}

\bibitem [\protect \citeauthoryear {%
Cornelissen%
, Peters%
\BCBL {}\ \BBA {} Palmer%
}{%
Cornelissen%
\ \protect \BOthers {.}}{%
{\protect \APACyear {2002}}%
}]{%
Cornelissen.BehavResMeth.2002}
\APACinsertmetastar {%
Cornelissen.BehavResMeth.2002}%
\begin{APACrefauthors}%
Cornelissen, F\BPBI W.%
, Peters, E\BPBI M.%
\BCBL {}\ \BBA {} Palmer, J.%
\end{APACrefauthors}%
\unskip\
\newblock
\APACrefYearMonthDay{2002}{}{}.
\newblock
{\BBOQ}\APACrefatitle {The {E}yelink {T}oolbox: {E}ye tracking with
  {M}{A}{T}{L}{A}{B} and the {P}sychophysics {T}oolbox} {The {E}yelink
  {T}oolbox: {E}ye tracking with {M}{A}{T}{L}{A}{B} and the {P}sychophysics
  {T}oolbox}.{\BBCQ}
\newblock
\APACjournalVolNumPages{Behavioral Research Methods, Instruments, \&
  Computers}{34}{4}{613--617}.
\newblock
\begin{APACrefDOI} \doi{10.3758/BF03195489} \end{APACrefDOI}
\PrintBackRefs{\CurrentBib}

\bibitem [\protect \citeauthoryear {%
Cousineau%
}{%
Cousineau%
}{%
{\protect \APACyear {2005}}%
}]{%
Cousineau.TQMP.2005}
\APACinsertmetastar {%
Cousineau.TQMP.2005}%
\begin{APACrefauthors}%
Cousineau, D.%
\end{APACrefauthors}%
\unskip\
\newblock
\APACrefYearMonthDay{2005}{}{}.
\newblock
{\BBOQ}\APACrefatitle {Confidence intervals in within-subject designs: {A}
  simpler solution to {L}oftus and {M}asson's method} {Confidence intervals in
  within-subject designs: {A} simpler solution to {L}oftus and {M}asson's
  method}.{\BBCQ}
\newblock
\APACjournalVolNumPages{Tutorial in Quantitative Methods for
  Psychology}{1}{1}{71--75}.
\newblock
\begin{APACrefURL} \url{http://tqmp.org/Content/vol01-1/p042/p042.pdf}
  \end{APACrefURL}
\PrintBackRefs{\CurrentBib}

\bibitem [\protect \citeauthoryear {%
Derrington%
\ \BBA {} Lennie%
}{%
Derrington%
\ \BBA {} Lennie%
}{%
{\protect \APACyear {1984}}%
}]{%
Derrington.JPhysiol.1984}
\APACinsertmetastar {%
Derrington.JPhysiol.1984}%
\begin{APACrefauthors}%
Derrington, A\BPBI M.%
\BCBT {}\ \BBA {} Lennie, P.%
\end{APACrefauthors}%
\unskip\
\newblock
\APACrefYearMonthDay{1984}{}{}.
\newblock
{\BBOQ}\APACrefatitle {Spatial and temporal contrast sensitivities of neurones
  in lateral geniculate nucleus of macaque} {Spatial and temporal contrast
  sensitivities of neurones in lateral geniculate nucleus of macaque}.{\BBCQ}
\newblock
\APACjournalVolNumPages{The Journal of Physiology}{357}{1}{219--240}.
\newblock
\begin{APACrefDOI} \doi{10.1113/jphysiol.1984.sp015498} \end{APACrefDOI}
\PrintBackRefs{\CurrentBib}

\bibitem [\protect \citeauthoryear {%
Engbert%
\ \BBA {} Kliegl%
}{%
Engbert%
\ \BBA {} Kliegl%
}{%
{\protect \APACyear {2003}}%
}]{%
Engbert.VisionRes.2003}
\APACinsertmetastar {%
Engbert.VisionRes.2003}%
\begin{APACrefauthors}%
Engbert, R.%
\BCBT {}\ \BBA {} Kliegl, R.%
\end{APACrefauthors}%
\unskip\
\newblock
\APACrefYearMonthDay{2003}{}{}.
\newblock
{\BBOQ}\APACrefatitle {Microsaccades uncover the orientation of covert
  attention} {Microsaccades uncover the orientation of covert
  attention}.{\BBCQ}
\newblock
\APACjournalVolNumPages{Vision Research}{43}{}{1035--1045}.
\newblock
\begin{APACrefDOI} \doi{10.1016/S0042-6989(03)00084-1} \end{APACrefDOI}
\PrintBackRefs{\CurrentBib}

\bibitem [\protect \citeauthoryear {%
Engbert%
\ \BBA {} Mergenthaler%
}{%
Engbert%
\ \BBA {} Mergenthaler%
}{%
{\protect \APACyear {2006}}%
}]{%
Engbert.PNAS.2006}
\APACinsertmetastar {%
Engbert.PNAS.2006}%
\begin{APACrefauthors}%
Engbert, R.%
\BCBT {}\ \BBA {} Mergenthaler, K.%
\end{APACrefauthors}%
\unskip\
\newblock
\APACrefYearMonthDay{2006}{}{}.
\newblock
{\BBOQ}\APACrefatitle {Microsaccades are triggered by low level retinal image
  slip} {Microsaccades are triggered by low level retinal image slip}.{\BBCQ}
\newblock
\APACjournalVolNumPages{Proceedings of the National Academy of Sciences of the
  United States of America}{103}{18}{7192--7197}.
\newblock
\begin{APACrefDOI} \doi{10.1073/pnas.0509557103} \end{APACrefDOI}
\PrintBackRefs{\CurrentBib}

\bibitem [\protect \citeauthoryear {%
Foulsham%
, Teszka%
\BCBL {}\ \BBA {} Kingstone%
}{%
Foulsham%
\ \protect \BOthers {.}}{%
{\protect \APACyear {2011}}%
}]{%
Foulsham.AttenPerceptPsychophys.2011}
\APACinsertmetastar {%
Foulsham.AttenPerceptPsychophys.2011}%
\begin{APACrefauthors}%
Foulsham, T.%
, Teszka, R.%
\BCBL {}\ \BBA {} Kingstone, A.%
\end{APACrefauthors}%
\unskip\
\newblock
\APACrefYearMonthDay{2011}{}{}.
\newblock
{\BBOQ}\APACrefatitle {Saccade control in natural images is shaped by the
  information visible at fixation: {E}vidence from asymmetric gaze-contingent
  windows} {Saccade control in natural images is shaped by the information
  visible at fixation: {E}vidence from asymmetric gaze-contingent
  windows}.{\BBCQ}
\newblock
\APACjournalVolNumPages{Attention, Perception, \&
  Psychophysics}{73}{}{266--283}.
\newblock
\begin{APACrefDOI} \doi{10.3758/s13414-010-0014-5} \end{APACrefDOI}
\PrintBackRefs{\CurrentBib}

\bibitem [\protect \citeauthoryear {%
Geringswald%
, Porracin%
\BCBL {}\ \BBA {} Pollmann%
}{%
Geringswald%
\ \protect \BOthers {.}}{%
{\protect \APACyear {2016}}%
}]{%
Geringswald.JVis.2016}
\APACinsertmetastar {%
Geringswald.JVis.2016}%
\begin{APACrefauthors}%
Geringswald, F.%
, Porracin, E.%
\BCBL {}\ \BBA {} Pollmann, S.%
\end{APACrefauthors}%
\unskip\
\newblock
\APACrefYearMonthDay{2016}{}{}.
\newblock
{\BBOQ}\APACrefatitle {Impairment of visual memory for objects in natural
  scenes by simulated central scotoma} {Impairment of visual memory for objects
  in natural scenes by simulated central scotoma}.{\BBCQ}
\newblock
\APACjournalVolNumPages{Journal of Vision}{16(2):6}{}{1--12}.
\newblock
\begin{APACrefDOI} \doi{10.1167/16.2.6} \end{APACrefDOI}
\PrintBackRefs{\CurrentBib}

\bibitem [\protect \citeauthoryear {%
Gilchrist%
}{%
Gilchrist%
}{%
{\protect \APACyear {2011}}%
}]{%
Gilchrist.INBOOK.2011}
\APACinsertmetastar {%
Gilchrist.INBOOK.2011}%
\begin{APACrefauthors}%
Gilchrist, I\BPBI D.%
\end{APACrefauthors}%
\unskip\
\newblock
\APACrefYearMonthDay{2011}{}{}.
\newblock
{\BBOQ}\APACrefatitle {Saccades} {Saccades}.{\BBCQ}
\newblock
\BIn{} S\BPBI P.~Liversedge, I\BPBI D.~Gilchrist\BCBL {}\ \BBA {} S.~Everling\
  (\BEDS), \APACrefbtitle {The {O}xford handbook of eye movements} {The
  {O}xford handbook of eye movements}\ (\BPGS\ 85--94).
\newblock
\APACaddressPublisher{}{Oxford: {O}xford {U}niversity {P}ress}.
\PrintBackRefs{\CurrentBib}

\bibitem [\protect \citeauthoryear {%
Hansen%
, Pracejus%
\BCBL {}\ \BBA {} Gegenfurtner%
}{%
Hansen%
\ \protect \BOthers {.}}{%
{\protect \APACyear {2009}}%
}]{%
Hansen.JVis.2009}
\APACinsertmetastar {%
Hansen.JVis.2009}%
\begin{APACrefauthors}%
Hansen, T.%
, Pracejus, L.%
\BCBL {}\ \BBA {} Gegenfurtner, K\BPBI R.%
\end{APACrefauthors}%
\unskip\
\newblock
\APACrefYearMonthDay{2009}{}{}.
\newblock
{\BBOQ}\APACrefatitle {Color perception in the intermediate periphery of the
  visual field} {Color perception in the intermediate periphery of the visual
  field}.{\BBCQ}
\newblock
\APACjournalVolNumPages{Journal of Vision}{9(4):26}{}{1--12}.
\newblock
\begin{APACrefDOI} \doi{10.1167/9.4.26} \end{APACrefDOI}
\PrintBackRefs{\CurrentBib}

\bibitem [\protect \citeauthoryear {%
Henderson%
\ \BBA {} Hollingworth%
}{%
Henderson%
\ \BBA {} Hollingworth%
}{%
{\protect \APACyear {1999}}%
}]{%
Henderson.PsycholSci.1999}
\APACinsertmetastar {%
Henderson.PsycholSci.1999}%
\begin{APACrefauthors}%
Henderson, J\BPBI M.%
\BCBT {}\ \BBA {} Hollingworth, A.%
\end{APACrefauthors}%
\unskip\
\newblock
\APACrefYearMonthDay{1999}{}{}.
\newblock
{\BBOQ}\APACrefatitle {The role of fixation position in detecting scene changes
  across saccades} {The role of fixation position in detecting scene changes
  across saccades}.{\BBCQ}
\newblock
\APACjournalVolNumPages{Psychological Science}{10}{5}{438--443}.
\newblock
\begin{APACrefDOI} \doi{10.1111/1467-9280.00183} \end{APACrefDOI}
\PrintBackRefs{\CurrentBib}

\bibitem [\protect \citeauthoryear {%
Henderson%
, Malcolm%
\BCBL {}\ \BBA {} Schandl%
}{%
Henderson%
\ \protect \BOthers {.}}{%
{\protect \APACyear {2009}}%
}]{%
Henderson.PsychonBullRev.2009}
\APACinsertmetastar {%
Henderson.PsychonBullRev.2009}%
\begin{APACrefauthors}%
Henderson, J\BPBI M.%
, Malcolm, G\BPBI L.%
\BCBL {}\ \BBA {} Schandl, C.%
\end{APACrefauthors}%
\unskip\
\newblock
\APACrefYearMonthDay{2009}{}{}.
\newblock
{\BBOQ}\APACrefatitle {Searching in the dark: {C}ognitive relevance drives
  attention in real-world scenes} {Searching in the dark: {C}ognitive relevance
  drives attention in real-world scenes}.{\BBCQ}
\newblock
\APACjournalVolNumPages{Psychonomic Bulletin \& Review}{16}{5}{850--856}.
\newblock
\begin{APACrefDOI} \doi{10.3758/PBR.16.5.850} \end{APACrefDOI}
\PrintBackRefs{\CurrentBib}

\bibitem [\protect \citeauthoryear {%
Henderson%
, McClure%
, Pierce%
\BCBL {}\ \BBA {} Schrock%
}{%
Henderson%
\ \protect \BOthers {.}}{%
{\protect \APACyear {1997}}%
}]{%
Henderson.PerceptPsychophys.1997}
\APACinsertmetastar {%
Henderson.PerceptPsychophys.1997}%
\begin{APACrefauthors}%
Henderson, J\BPBI M.%
, McClure, K\BPBI K.%
, Pierce, S.%
\BCBL {}\ \BBA {} Schrock, G.%
\end{APACrefauthors}%
\unskip\
\newblock
\APACrefYearMonthDay{1997}{}{}.
\newblock
{\BBOQ}\APACrefatitle {Object identification without foveal vision: {E}vidence
  from an artificial scotoma paradigm} {Object identification without foveal
  vision: {E}vidence from an artificial scotoma paradigm}.{\BBCQ}
\newblock
\APACjournalVolNumPages{Perception \& Psychophysics}{59}{3}{323--346}.
\newblock
\begin{APACrefDOI} \doi{10.3758/BF03211901} \end{APACrefDOI}
\PrintBackRefs{\CurrentBib}

\bibitem [\protect \citeauthoryear {%
Henderson%
, Williams%
, Castelhano%
\BCBL {}\ \BBA {} Falk%
}{%
Henderson%
\ \protect \BOthers {.}}{%
{\protect \APACyear {2003}}%
}]{%
Henderson.PerceptPsychophys.2003}
\APACinsertmetastar {%
Henderson.PerceptPsychophys.2003}%
\begin{APACrefauthors}%
Henderson, J\BPBI M.%
, Williams, C\BPBI C.%
, Castelhano, M\BPBI S.%
\BCBL {}\ \BBA {} Falk, R\BPBI J.%
\end{APACrefauthors}%
\unskip\
\newblock
\APACrefYearMonthDay{2003}{}{}.
\newblock
{\BBOQ}\APACrefatitle {Eye movements and picture processing during recognition}
  {Eye movements and picture processing during recognition}.{\BBCQ}
\newblock
\APACjournalVolNumPages{Perception \& Psychophysics}{65}{5}{725--734}.
\newblock
\begin{APACrefDOI} \doi{10.3758/BF03194809} \end{APACrefDOI}
\PrintBackRefs{\CurrentBib}

\bibitem [\protect \citeauthoryear {%
Hilz%
\ \BBA {} Cavonius%
}{%
Hilz%
\ \BBA {} Cavonius%
}{%
{\protect \APACyear {1974}}%
}]{%
Hilz.VisionRes.1974}
\APACinsertmetastar {%
Hilz.VisionRes.1974}%
\begin{APACrefauthors}%
Hilz, R.%
\BCBT {}\ \BBA {} Cavonius, C\BPBI R.%
\end{APACrefauthors}%
\unskip\
\newblock
\APACrefYearMonthDay{1974}{}{}.
\newblock
{\BBOQ}\APACrefatitle {Functional organization of the peripheral retina:
  {S}ensitivity to periodic stimuli} {Functional organization of the peripheral
  retina: {S}ensitivity to periodic stimuli}.{\BBCQ}
\newblock
\APACjournalVolNumPages{Vision Research}{14}{}{1333--1337}.
\newblock
\begin{APACrefDOI} \doi{10.1016/0042-6989(74)90006-6} \end{APACrefDOI}
\PrintBackRefs{\CurrentBib}

\bibitem [\protect \citeauthoryear {%
Hwang%
, Higgins%
\BCBL {}\ \BBA {} Pomplun%
}{%
Hwang%
\ \protect \BOthers {.}}{%
{\protect \APACyear {2007}}%
}]{%
Hwang.Proceedings.2007}
\APACinsertmetastar {%
Hwang.Proceedings.2007}%
\begin{APACrefauthors}%
Hwang, A\BPBI D.%
, Higgins, E\BPBI C.%
\BCBL {}\ \BBA {} Pomplun, M.%
\end{APACrefauthors}%
\unskip\
\newblock
\APACrefYearMonthDay{2007}{}{}.
\newblock
{\BBOQ}\APACrefatitle {How chromaticity guides visual search in real-world
  scenes} {How chromaticity guides visual search in real-world scenes}.{\BBCQ}
\newblock
\BIn{} \APACrefbtitle {Proceedings of the {A}nnual {M}eeting of the {C}ognitive
  {S}cience {S}ociety} {Proceedings of the {A}nnual {M}eeting of the
  {C}ognitive {S}cience {S}ociety}\ (\BVOL~29).
\PrintBackRefs{\CurrentBib}

\bibitem [\protect \citeauthoryear {%
Hwang%
, Higgins%
\BCBL {}\ \BBA {} Pomplun%
}{%
Hwang%
\ \protect \BOthers {.}}{%
{\protect \APACyear {2009}}%
}]{%
Hwang.JVis.2009}
\APACinsertmetastar {%
Hwang.JVis.2009}%
\begin{APACrefauthors}%
Hwang, A\BPBI D.%
, Higgins, E\BPBI C.%
\BCBL {}\ \BBA {} Pomplun, M.%
\end{APACrefauthors}%
\unskip\
\newblock
\APACrefYearMonthDay{2009}{}{}.
\newblock
{\BBOQ}\APACrefatitle {A model of top-down attentional control during visual
  search in complex scenes} {A model of top-down attentional control during
  visual search in complex scenes}.{\BBCQ}
\newblock
\APACjournalVolNumPages{Journal of Vision}{9(5):25}{}{1--18}.
\newblock
\begin{APACrefDOI} \doi{10.1167/9.5.25} \end{APACrefDOI}
\PrintBackRefs{\CurrentBib}

\bibitem [\protect \citeauthoryear {%
Johnson%
}{%
Johnson%
}{%
{\protect \APACyear {1986}}%
}]{%
Johnson.AmJOptomPhysiolOptics.1986}
\APACinsertmetastar {%
Johnson.AmJOptomPhysiolOptics.1986}%
\begin{APACrefauthors}%
Johnson, M\BPBI A.%
\end{APACrefauthors}%
\unskip\
\newblock
\APACrefYearMonthDay{1986}{}{}.
\newblock
{\BBOQ}\APACrefatitle {Color vision in the peripheral retina} {Color vision in
  the peripheral retina}.{\BBCQ}
\newblock
\APACjournalVolNumPages{American Journal of Optometry \& Physiological
  Optics}{63}{2}{97--103}.
\newblock
\begin{APACrefDOI} \doi{10.1097/00006324-198602000-00003} \end{APACrefDOI}
\PrintBackRefs{\CurrentBib}

\bibitem [\protect \citeauthoryear {%
Jones%
\ \BBA {} Higgins%
}{%
Jones%
\ \BBA {} Higgins%
}{%
{\protect \APACyear {1947}}%
}]{%
Jones.JOptSocAm.1947}
\APACinsertmetastar {%
Jones.JOptSocAm.1947}%
\begin{APACrefauthors}%
Jones, L\BPBI A.%
\BCBT {}\ \BBA {} Higgins, G\BPBI C.%
\end{APACrefauthors}%
\unskip\
\newblock
\APACrefYearMonthDay{1947}{}{}.
\newblock
{\BBOQ}\APACrefatitle {Photographic granularity and graininess. {I}{I}{I}.
  {S}ome characteristics of the visual system of importance in the evaluation
  of graininess and granularity} {Photographic granularity and graininess.
  {I}{I}{I}. {S}ome characteristics of the visual system of importance in the
  evaluation of graininess and granularity}.{\BBCQ}
\newblock
\APACjournalVolNumPages{Journal of the Optical Society of
  America}{37}{4}{217--263}.
\newblock
\begin{APACrefDOI} \doi{10.1364/JOSA.37.000217} \end{APACrefDOI}
\PrintBackRefs{\CurrentBib}

\bibitem [\protect \citeauthoryear {%
Kleiner%
, Brainard%
\BCBL {}\ \BBA {} Pelli%
}{%
Kleiner%
\ \protect \BOthers {.}}{%
{\protect \APACyear {2007}}%
}]{%
Kleiner.Perception.2007}
\APACinsertmetastar {%
Kleiner.Perception.2007}%
\begin{APACrefauthors}%
Kleiner, M.%
, Brainard, D\BPBI H.%
\BCBL {}\ \BBA {} Pelli, D\BPBI G.%
\end{APACrefauthors}%
\unskip\
\newblock
\APACrefYearMonthDay{2007}{}{}.
\newblock
{\BBOQ}\APACrefatitle {What's new in {P}sychtoolbox-3?} {What's new in
  {P}sychtoolbox-3?}{\BBCQ}
\newblock
\APACjournalVolNumPages{Perception}{36}{}{14}.
\PrintBackRefs{\CurrentBib}

\bibitem [\protect \citeauthoryear {%
Larson%
\ \BBA {} Loschky%
}{%
Larson%
\ \BBA {} Loschky%
}{%
{\protect \APACyear {2009}}%
}]{%
Larson.JVis.2009}
\APACinsertmetastar {%
Larson.JVis.2009}%
\begin{APACrefauthors}%
Larson, A\BPBI M.%
\BCBT {}\ \BBA {} Loschky, L\BPBI C.%
\end{APACrefauthors}%
\unskip\
\newblock
\APACrefYearMonthDay{2009}{}{}.
\newblock
{\BBOQ}\APACrefatitle {The contributions of central versus peripheral vision to
  scene gist recognition} {The contributions of central versus peripheral
  vision to scene gist recognition}.{\BBCQ}
\newblock
\APACjournalVolNumPages{Journal of Vision}{9(10):6}{}{1--16}.
\newblock
\begin{APACrefDOI} \doi{10.1167/9.10.6} \end{APACrefDOI}
\PrintBackRefs{\CurrentBib}

\bibitem [\protect \citeauthoryear {%
Laubrock%
, Cajar%
\BCBL {}\ \BBA {} Engbert%
}{%
Laubrock%
\ \protect \BOthers {.}}{%
{\protect \APACyear {2013}}%
}]{%
Laubrock.JVis.2013}
\APACinsertmetastar {%
Laubrock.JVis.2013}%
\begin{APACrefauthors}%
Laubrock, J.%
, Cajar, A.%
\BCBL {}\ \BBA {} Engbert, R.%
\end{APACrefauthors}%
\unskip\
\newblock
\APACrefYearMonthDay{2013}{}{}.
\newblock
{\BBOQ}\APACrefatitle {Control of fixation duration during scene viewing by
  interaction of foveal and peripheral processing} {Control of fixation
  duration during scene viewing by interaction of foveal and peripheral
  processing}.{\BBCQ}
\newblock
\APACjournalVolNumPages{Journal of Vision}{13(12):11}{}{1--20}.
\newblock
\begin{APACrefDOI} \doi{10.1167/13.12.11} \end{APACrefDOI}
\PrintBackRefs{\CurrentBib}

\bibitem [\protect \citeauthoryear {%
Loschky%
\ \BBA {} McConkie%
}{%
Loschky%
\ \BBA {} McConkie%
}{%
{\protect \APACyear {2002}}%
}]{%
Loschky.JExpPsycholAppl.2002}
\APACinsertmetastar {%
Loschky.JExpPsycholAppl.2002}%
\begin{APACrefauthors}%
Loschky, L\BPBI C.%
\BCBT {}\ \BBA {} McConkie, G\BPBI W.%
\end{APACrefauthors}%
\unskip\
\newblock
\APACrefYearMonthDay{2002}{}{}.
\newblock
{\BBOQ}\APACrefatitle {Investigating spatial vision and dynamic attentional
  selection using a gaze-contingent multiresolutional display} {Investigating
  spatial vision and dynamic attentional selection using a gaze-contingent
  multiresolutional display}.{\BBCQ}
\newblock
\APACjournalVolNumPages{Journal of Experimental Psychology:
  Applied}{8}{2}{99--117}.
\newblock
\begin{APACrefDOI} \doi{10.1037//1076-898X.8.2.99} \end{APACrefDOI}
\PrintBackRefs{\CurrentBib}

\bibitem [\protect \citeauthoryear {%
Loschky%
, McConkie%
, Yang%
\BCBL {}\ \BBA {} Miller%
}{%
Loschky%
\ \protect \BOthers {.}}{%
{\protect \APACyear {2005}}%
}]{%
Loschky.VisCogn.2005}
\APACinsertmetastar {%
Loschky.VisCogn.2005}%
\begin{APACrefauthors}%
Loschky, L\BPBI C.%
, McConkie, G\BPBI W.%
, Yang, J.%
\BCBL {}\ \BBA {} Miller, M\BPBI E.%
\end{APACrefauthors}%
\unskip\
\newblock
\APACrefYearMonthDay{2005}{}{}.
\newblock
{\BBOQ}\APACrefatitle {The limits of visual resolution in natural scene
  viewing} {The limits of visual resolution in natural scene viewing}.{\BBCQ}
\newblock
\APACjournalVolNumPages{Visual Cognition}{12}{6}{1057--1092}.
\newblock
\begin{APACrefDOI} \doi{10.1080/13506280444000652} \end{APACrefDOI}
\PrintBackRefs{\CurrentBib}

\bibitem [\protect \citeauthoryear {%
Malcolm%
\ \BBA {} Henderson%
}{%
Malcolm%
\ \BBA {} Henderson%
}{%
{\protect \APACyear {2009}}%
}]{%
Malcolm.JVis.2009}
\APACinsertmetastar {%
Malcolm.JVis.2009}%
\begin{APACrefauthors}%
Malcolm, G\BPBI L.%
\BCBT {}\ \BBA {} Henderson, J\BPBI M.%
\end{APACrefauthors}%
\unskip\
\newblock
\APACrefYearMonthDay{2009}{}{}.
\newblock
{\BBOQ}\APACrefatitle {The effects of target template specificity on visual
  search in real-world scenes: {E}vidence from eye movements} {The effects of
  target template specificity on visual search in real-world scenes: {E}vidence
  from eye movements}.{\BBCQ}
\newblock
\APACjournalVolNumPages{Journal of Vision}{9(11):8}{}{1--13}.
\newblock
\begin{APACrefDOI} \doi{10.1167/9.11.8} \end{APACrefDOI}
\PrintBackRefs{\CurrentBib}

\bibitem [\protect \citeauthoryear {%
Malcolm%
\ \BBA {} Henderson%
}{%
Malcolm%
\ \BBA {} Henderson%
}{%
{\protect \APACyear {2010}}%
}]{%
Malcolm.JVis.2010}
\APACinsertmetastar {%
Malcolm.JVis.2010}%
\begin{APACrefauthors}%
Malcolm, G\BPBI L.%
\BCBT {}\ \BBA {} Henderson, J\BPBI M.%
\end{APACrefauthors}%
\unskip\
\newblock
\APACrefYearMonthDay{2010}{}{}.
\newblock
{\BBOQ}\APACrefatitle {Combining top-down processes to guide eye movements
  during real-world scene search} {Combining top-down processes to guide eye
  movements during real-world scene search}.{\BBCQ}
\newblock
\APACjournalVolNumPages{Journal of Vision}{10(2):4}{}{1--11}.
\newblock
\begin{APACrefDOI} \doi{10.1167/10.2.4} \end{APACrefDOI}
\PrintBackRefs{\CurrentBib}

\bibitem [\protect \citeauthoryear {%
Mohr%
\ \protect \BOthers {.}}{%
Mohr%
\ \protect \BOthers {.}}{%
{\protect \APACyear {2016}}%
}]{%
Mohr.FrontiersPsychol.2016}
\APACinsertmetastar {%
Mohr.FrontiersPsychol.2016}%
\begin{APACrefauthors}%
Mohr, J.%
, Seyfarth, J.%
, Lueschow, A.%
, Weber, J\BPBI E.%
, Wichmann, F\BPBI A.%
\BCBL {}\ \BBA {} Obermayer, K.%
\end{APACrefauthors}%
\unskip\
\newblock
\APACrefYearMonthDay{2016}{}{}.
\newblock
{\BBOQ}\APACrefatitle {{B}{O}i{S}--{B}erlin {O}bject in {S}cene Database:
  Controlled Photographic Images for Visual Search Experiments with Quantified
  Contextual Priors} {{B}{O}i{S}--{B}erlin {O}bject in {S}cene database:
  Controlled photographic images for visual search experiments with quantified
  contextual priors}.{\BBCQ}
\newblock
\APACjournalVolNumPages{Frontiers in Psychology}{7}{}{749}.
\newblock
\begin{APACrefDOI} \doi{10.3389/fpsyg.2016.00749} \end{APACrefDOI}
\PrintBackRefs{\CurrentBib}

\bibitem [\protect \citeauthoryear {%
Morey%
}{%
Morey%
}{%
{\protect \APACyear {2008}}%
}]{%
Morey.TQMP.2008}
\APACinsertmetastar {%
Morey.TQMP.2008}%
\begin{APACrefauthors}%
Morey, R\BPBI D.%
\end{APACrefauthors}%
\unskip\
\newblock
\APACrefYearMonthDay{2008}{}{}.
\newblock
{\BBOQ}\APACrefatitle {Confidence intervals from normalized data: {A}
  correction to {C}ousineau (2005)} {Confidence intervals from normalized data:
  {A} correction to {C}ousineau (2005)}.{\BBCQ}
\newblock
\APACjournalVolNumPages{Tutorial in Quantitative Methods for
  Psychology}{42}{2}{61--64}.
\newblock
\begin{APACrefURL}
  \url{http://pcl.missouri.edu/sites/default/files/morey.2008.pdf}
  \end{APACrefURL}
\PrintBackRefs{\CurrentBib}

\bibitem [\protect \citeauthoryear {%
Neider%
\ \BBA {} Zelinsky%
}{%
Neider%
\ \BBA {} Zelinsky%
}{%
{\protect \APACyear {2006}}%
}]{%
Neider.VisionRes.2006}
\APACinsertmetastar {%
Neider.VisionRes.2006}%
\begin{APACrefauthors}%
Neider, M\BPBI B.%
\BCBT {}\ \BBA {} Zelinsky, G\BPBI J.%
\end{APACrefauthors}%
\unskip\
\newblock
\APACrefYearMonthDay{2006}{}{}.
\newblock
{\BBOQ}\APACrefatitle {Scene context guides eye movements during visual search}
  {Scene context guides eye movements during visual search}.{\BBCQ}
\newblock
\APACjournalVolNumPages{Vision Research}{10}{5}{614--621}.
\newblock
\begin{APACrefDOI} \doi{10.1016/j.visres.2005.08.025} \end{APACrefDOI}
\PrintBackRefs{\CurrentBib}

\bibitem [\protect \citeauthoryear {%
Nuthmann%
}{%
Nuthmann%
}{%
{\protect \APACyear {2013}}%
}]{%
Nuthmann.VisCogn.2013}
\APACinsertmetastar {%
Nuthmann.VisCogn.2013}%
\begin{APACrefauthors}%
Nuthmann, A.%
\end{APACrefauthors}%
\unskip\
\newblock
\APACrefYearMonthDay{2013}{}{}.
\newblock
{\BBOQ}\APACrefatitle {On the visual span during object search in real-world
  scenes} {On the visual span during object search in real-world
  scenes}.{\BBCQ}
\newblock
\APACjournalVolNumPages{Visual Cognition}{21}{7}{803--837}.
\newblock
\begin{APACrefDOI} \doi{10.1080/13506285.2013.832449} \end{APACrefDOI}
\PrintBackRefs{\CurrentBib}

\bibitem [\protect \citeauthoryear {%
Nuthmann%
}{%
Nuthmann%
}{%
{\protect \APACyear {2014}}%
}]{%
Nuthmann.JExpPsycholHuman.2014}
\APACinsertmetastar {%
Nuthmann.JExpPsycholHuman.2014}%
\begin{APACrefauthors}%
Nuthmann, A.%
\end{APACrefauthors}%
\unskip\
\newblock
\APACrefYearMonthDay{2014}{}{}.
\newblock
{\BBOQ}\APACrefatitle {How do the regions of the visual field contribute to
  object search in real-world scenes? {E}vidence from eye movements} {How do
  the regions of the visual field contribute to object search in real-world
  scenes? {E}vidence from eye movements}.{\BBCQ}
\newblock
\APACjournalVolNumPages{Journal of Experimental Psychology: Human Perception
  and Performance}{40}{1}{342--360}.
\newblock
\begin{APACrefDOI} \doi{10.1037/a0033854} \end{APACrefDOI}
\PrintBackRefs{\CurrentBib}

\bibitem [\protect \citeauthoryear {%
Nuthmann%
\ \BBA {} Malcolm%
}{%
Nuthmann%
\ \BBA {} Malcolm%
}{%
{\protect \APACyear {2016}}%
}]{%
Nuthmann.JVis.2016}
\APACinsertmetastar {%
Nuthmann.JVis.2016}%
\begin{APACrefauthors}%
Nuthmann, A.%
\BCBT {}\ \BBA {} Malcolm, G\BPBI L.%
\end{APACrefauthors}%
\unskip\
\newblock
\APACrefYearMonthDay{2016}{}{}.
\newblock
{\BBOQ}\APACrefatitle {Eye-guidance during real-world scene search: {T}he role
  color plays in central and peripheral vision} {Eye-guidance during real-world
  scene search: {T}he role color plays in central and peripheral
  vision}.{\BBCQ}
\newblock
\APACjournalVolNumPages{Journal of Vision}{16(2):3}{}{1--16}.
\newblock
\begin{APACrefDOI} \doi{10.1167/16.2.3} \end{APACrefDOI}
\PrintBackRefs{\CurrentBib}

\bibitem [\protect \citeauthoryear {%
Peyrin%
, Chauvin%
, Chokron%
\BCBL {}\ \BBA {} Marendaz%
}{%
Peyrin%
\ \protect \BOthers {.}}{%
{\protect \APACyear {2003}}%
}]{%
Peyrin.BrainCogn.2003}
\APACinsertmetastar {%
Peyrin.BrainCogn.2003}%
\begin{APACrefauthors}%
Peyrin, C.%
, Chauvin, A.%
, Chokron, S.%
\BCBL {}\ \BBA {} Marendaz, C.%
\end{APACrefauthors}%
\unskip\
\newblock
\APACrefYearMonthDay{2003}{}{}.
\newblock
{\BBOQ}\APACrefatitle {Hemispheric specialization for spatial frequency
  processing in the analysis of natural scenes} {Hemispheric specialization for
  spatial frequency processing in the analysis of natural scenes}.{\BBCQ}
\newblock
\APACjournalVolNumPages{Brain and Cognition}{53}{}{278--282}.
\newblock
\begin{APACrefDOI} \doi{10.1016/S0278-2626(03)00126-X} \end{APACrefDOI}
\PrintBackRefs{\CurrentBib}

\bibitem [\protect \citeauthoryear {%
{R Core Team}%
}{%
{R Core Team}%
}{%
{\protect \APACyear {2018}}%
}]{%
RCoreTeam.2018}
\APACinsertmetastar {%
RCoreTeam.2018}%
\begin{APACrefauthors}%
{R Core Team}.%
\end{APACrefauthors}%
\unskip\
\newblock
\APACrefYearMonthDay{2018}{}{}.
\newblock
{\BBOQ}\APACrefatitle {R: {A} language and environment for statistical
  computing} {R: {A} language and environment for statistical
  computing}{\BBCQ}\ [\bibcomputersoftwaremanual].
\newblock
\APACaddressPublisher{}{Vienna, Austria}.
\newblock
\begin{APACrefURL} \url{https://www.R-project.org/} \end{APACrefURL}
\PrintBackRefs{\CurrentBib}

\bibitem [\protect \citeauthoryear {%
Spotorno%
, Malcolm%
\BCBL {}\ \BBA {} Tatler%
}{%
Spotorno%
\ \protect \BOthers {.}}{%
{\protect \APACyear {2014}}%
}]{%
Spotorno.JVis.2014}
\APACinsertmetastar {%
Spotorno.JVis.2014}%
\begin{APACrefauthors}%
Spotorno, S.%
, Malcolm, G\BPBI L.%
\BCBL {}\ \BBA {} Tatler, B\BPBI W.%
\end{APACrefauthors}%
\unskip\
\newblock
\APACrefYearMonthDay{2014}{}{}.
\newblock
{\BBOQ}\APACrefatitle {How context information and target information guide the
  eyes from the first epoch of search in real-world scenes} {How context
  information and target information guide the eyes from the first epoch of
  search in real-world scenes}.{\BBCQ}
\newblock
\APACjournalVolNumPages{Journal of Vision}{14(2):7}{}{1--21}.
\newblock
\begin{APACrefDOI} \doi{10.1167/14.2.7} \end{APACrefDOI}
\PrintBackRefs{\CurrentBib}

\bibitem [\protect \citeauthoryear {%
Tatler%
, Hayhoe%
, Land%
\BCBL {}\ \BBA {} Ballard%
}{%
Tatler%
\ \protect \BOthers {.}}{%
{\protect \APACyear {2011}}%
}]{%
Tatler.JVis.2011}
\APACinsertmetastar {%
Tatler.JVis.2011}%
\begin{APACrefauthors}%
Tatler, B\BPBI W.%
, Hayhoe, M\BPBI M.%
, Land, M\BPBI F.%
\BCBL {}\ \BBA {} Ballard, D\BPBI H.%
\end{APACrefauthors}%
\unskip\
\newblock
\APACrefYearMonthDay{2011}{}{}.
\newblock
{\BBOQ}\APACrefatitle {Eye guidance in natural vision: {R}einterpreting
  salience} {Eye guidance in natural vision: {R}einterpreting salience}.{\BBCQ}
\newblock
\APACjournalVolNumPages{Journal of Vision}{11(5):5}{}{1--23}.
\newblock
\begin{APACrefDOI} \doi{10.1167/11.5.5} \end{APACrefDOI}
\PrintBackRefs{\CurrentBib}

\bibitem [\protect \citeauthoryear {%
{van Diepen}%
, Wampers%
\BCBL {}\ \BBA {} d'Ydewalle%
}{%
{van Diepen}%
\ \protect \BOthers {.}}{%
{\protect \APACyear {1998}}%
}]{%
vanDiepen.INBOOK.1998}
\APACinsertmetastar {%
vanDiepen.INBOOK.1998}%
\begin{APACrefauthors}%
{van Diepen}, P\BPBI M\BPBI J.%
, Wampers, M.%
\BCBL {}\ \BBA {} d'Ydewalle, G.%
\end{APACrefauthors}%
\unskip\
\newblock
\APACrefYearMonthDay{1998}{}{}.
\newblock
{\BBOQ}\APACrefatitle {Eye guidance in reading and scene perception} {Eye
  guidance in reading and scene perception}.{\BBCQ}
\newblock
\BIn{} G\BPBI D\BPBI M.~Underwood\ (\BED), (\BPGS\ 337--355).
\newblock
\APACaddressPublisher{}{Oxford, {UK}: {E}lsevier}.
\PrintBackRefs{\CurrentBib}

\bibitem [\protect \citeauthoryear {%
Venables%
\ \BBA {} Ripley%
}{%
Venables%
\ \BBA {} Ripley%
}{%
{\protect \APACyear {2002}}%
}]{%
Venables.MASS.2002}
\APACinsertmetastar {%
Venables.MASS.2002}%
\begin{APACrefauthors}%
Venables, W\BPBI N.%
\BCBT {}\ \BBA {} Ripley, B\BPBI D.%
\end{APACrefauthors}%
\unskip\
\newblock
\APACrefYear{2002}.
\newblock
\APACrefbtitle {Modern applied statistics with {S}} {Modern applied statistics
  with {S}}.
\newblock
\APACaddressPublisher{}{New {Y}ork: {S}pringer}.
\PrintBackRefs{\CurrentBib}

\bibitem [\protect \citeauthoryear {%
Wells-Gray%
, Choi%
, Bries%
\BCBL {}\ \BBA {} Doble%
}{%
Wells-Gray%
\ \protect \BOthers {.}}{%
{\protect \APACyear {2016}}%
}]{%
Wells-Gray.Eye.2016}
\APACinsertmetastar {%
Wells-Gray.Eye.2016}%
\begin{APACrefauthors}%
Wells-Gray, E\BPBI M.%
, Choi, S\BPBI S.%
, Bries, A.%
\BCBL {}\ \BBA {} Doble, N.%
\end{APACrefauthors}%
\unskip\
\newblock
\APACrefYearMonthDay{2016}{}{}.
\newblock
{\BBOQ}\APACrefatitle {Variation in rod and cone density from the fovea to the
  mid-periphery in healthy human retinas using adaptive optics scanning laser
  ophtalmoscopy} {Variation in rod and cone density from the fovea to the
  mid-periphery in healthy human retinas using adaptive optics scanning laser
  ophtalmoscopy}.{\BBCQ}
\newblock
\APACjournalVolNumPages{Eye}{30}{}{1135--1143}.
\newblock
\begin{APACrefDOI} \doi{10.1038/eye.2016.107} \end{APACrefDOI}
\PrintBackRefs{\CurrentBib}

\bibitem [\protect \citeauthoryear {%
Wertheim%
}{%
Wertheim%
}{%
{\protect \APACyear {1894}}%
}]{%
Wertheim.ZPsycholPhysiolSinnesorgane.1894}
\APACinsertmetastar {%
Wertheim.ZPsycholPhysiolSinnesorgane.1894}%
\begin{APACrefauthors}%
Wertheim, T.%
\end{APACrefauthors}%
\unskip\
\newblock
\APACrefYearMonthDay{1894}{}{}.
\newblock
{\BBOQ}\APACrefatitle {{\"U}ber die indirekte {S}ehsch\"arfe} {{\"U}ber die
  indirekte {S}ehsch\"arfe}.{\BBCQ}
\newblock
\APACjournalVolNumPages{Zeitschrift f\"ur {P}sychologie und {P}hysiologie der
  {S}innesorgane}{7}{}{121--187}.
\PrintBackRefs{\CurrentBib}

\end{thebibliography}

\end{document}